\documentclass[prc,twocolumn,secnumroman,tightenlines,amssymb,aps,nobibnotes,
               superscriptaddress,showpacs,balancelastpage,floatfix,nofootinbib]{revtex4}
\usepackage{graphicx}
\usepackage{multirow}           
\usepackage{color}
\expandafter\ifx\csname package@font\endcsname\relax\else
\expandafter\expandafter
\expandafter\usepackage
\expandafter{\csname package@font\endcsname}
\fi
\DeclareRobustCommand\substyle{\name@idx{document substyle}}
\DeclareRobustCommand\classoption{\name@idx{document class option}}
\DeclareRobustCommand\classname{\name@idx{document class}}
\def\name@idx#1#2{{\ttfamily#2}
\index{#2\space#1=\string\ttt{#2}\space#1}\index{#1>#2=\string\ttt{#2}}}
\begin{document}

% Without \draft below the PACS numbers do not appear

%\draft

%\preprint{\fbox{\sc version of \today}}

%\def\lt{\raisebox{0.2ex}{$<$}
%\def\gt{\raisebox{0.2ex}{$>$}}

% The open-square-bracket below is closed before the \end{document}

%\twocolumn[\columnwidth\textwidth\csname@twocolumnfalse\endcsname

\title[Shape-dependent congruence energy]
      {Dynamical evolution of the spectator systems produced in\\
ultra-relativistic heavy-ion collisions}
\author{K. Mazurek}
\email{Katarzyna.Mazurek@ifj.edu.pl}
\affiliation{Institute of Nuclear Physics PAN, ul.\,Radzikowskiego 152,
             Pl-31342 Krak\'ow, Poland}
             
\author{A. Szczurek}
\email{Antoni.Szczurek@ifj.edu.pl}
\affiliation{Institute of Nuclear Physics PAN, ul.\,Radzikowskiego 152,
             Pl-31342 Krak\'ow, Poland}

\author{C. Schmitt}
\email{christelle.schmitt@iphc.cnrs.fr }
\affiliation{Institut Pluridisciplinaire Hubert Curien, 23 rue du Loess, B.P.\,28, F-67037 Strasbourg Cedex 2, France}
             
\author{P.N. Nadtochy}
\email{nadtoch77@gmail.com}
\affiliation{Omsk State Technical University, Mira prospekt 11, Omsk, 644050, Russia}

\date{\today}

\begin{abstract}
In peripheral heavy-ion collisions at ultra-relativistic energies,
usually only parts of the colliding nuclei 
effectively interact with each other. In the overlapping zone, a fireball or quark-gluon plasma is produced. The excitation 
energy of the heavy remnant can range from a few tens to several hundreds of MeV, depending on the impact parameter. The decay of these excited spectators 
is investigated in this work for the first time within a dynamical approach based on the multi-dimensional 
stochastic Langevin equation. The potential of this exploratory work to understand the connection between electromagnetic fields generated by the heavy 
spectators and measured pion distributions is discussed.
\end{abstract}

%Uncomment for PACS numbers title message
% Keywords required only for MST, PB, PMB, PM, JOA, JOB? 
%\vspace{2pc}
%\noindent{\it Keywords}: Article preparation, IOP journals
% Uncomment for Submitted to journal title message
%\submitto{\JPA}
% Comment out if separate title page not required

\maketitle

%%%%%%%%%%%%%%%%%%%%%%%%%%%%%%%%%%%%%%%%%%%%%%%%%%%%%%%%%%%%%%%%%%%%%%%%%%%%%%%%
%%%%%%%%%%%%%%%%%%%%%%%%%%%%%%%%%%%%%%%%%%%%%%%%%%%%%%%%%%%%%%%%%%%%%%%%%%%%%%%%
%%%%%%%%%%%%%%%%%%%%%%%%%%%%%%%%%%%%%%%%%%%%%%%%%%%%%%%%%%%%%%%%%%%%%%%%%%%%%%
%%%%%%%%%%%%%%%%%%%%%%%%%%%%%%%%%%%%%%%%%%%%%%%%%%%%%%%%%%%%%%%%%%%%%%%%%%%%%%%%

\section{Introduction}
\label{Sect.I}
%%%%%%%%%%%%%%%%%%%%%%%%%%%%%%%%%%%%%%%%%%%%%%%%%%%%%%%%%%%%%%%%%%%%%%%%%%%%%
%%%%%%%%%%%%%%%%%%%%%%%%%%%%%%%%%%%%%%%%%%%%%%%%%%%%%%%%%%%%%%%%%%%%%%%%%%%%%%%%

%\section*{References}

%\begin{widetext}
Collisions between heavy ions flying with ultra-relativistic velocities are studied theoretically and experimentally since many years, with the main goal being the  study of the properties of the quark-gluon plasma (QGP). In the present work, the attention is devoted to the properties and decay of the heavy remnants of the collision.
Non-central collisions unambiguously lead to azimuthal asymmetries in the pion trajectory \cite{rybicki:2014}, which 
may be linked to the electromagnetic field generated by these fast-moving charged remnants. The influence of this field on 
charged pions was discussed in \cite{rybicki:2007,rybicki:2013}.\\

Previous work confirmed that the collision between two heavy ions at (ultra-)relativistic energy can be viewed as a two-step process.  
The first stage of the collision, often referred to as abrasion, is a
very fast process. 
The two remnants of the collision are 
considered as 'spectators': They are characterized by some mass deficit as compared to the mass of the reaction partners, but 
follow their initial path almost undisturbed. In the (participant) collision zone, at typical CERN  Super Proton Synchrotron 
(SPS) energies, pions or other hadrons are produced and form a quark-gluon plasma. The mass of the spectator remnants, and 
accordingly the number of 
nucleons involved in the fireball, depend on the impact parameter. 
The second stage of the collision, in 
comparison to the first one, 
is a slow process. Along this stage, often referred to as ablation, 
the primary hot products release their excitation energy and 
decay to a stable state by emitting light particles and $\gamma$-rays. Heavy spectators have additionally a large probability 
to decay by fission. During this time, the fireball expands and produced
particles (mostly pions) which fly apart. 

The fireball was discussed in the context of the abrasion model already in
\cite{bowman:1973,morrissey:1978,swiatecki:1976}. A new calculation \cite{szczurek:2017}, based on simple 
Coulomb effects, suggests that at SPS energies  
the QGP plasma creates a kind of fire-streaks, which velocity along the collision axis changes across the impact parameter space. It 
is assumed that the spectators do not feel the interaction with the plasma. Also, it is believed that the decay of the spectators is 
unimportant for the evolution of the QGP and subsequent hadronization. However, the strong electromagnetic field generated by fast-moving 
spectators can act {\it e.g.} on the charged pions created from the fire-streaks of the QGP. Electromagnetic effects 
yield different distortions of the
positively and negatively charged pions. They lead to a damping of $\pi^+$ and an enhancement of $\pi^-$ for pions moving with 
velocity equal to the velocity of the spectators. Long-range electromagnetic interactions \cite{rybicki:2007} are possible provided 
 that spectators live long enough. Consequently, a realistic estimate of this time, and understanding of what happens not 
 only to the plasma but also to the spectators, seems interesting and important in this context.\\

While previous works focused on the plasma, the present study is
dedicated to the properties and decay of the heavy spectators. 
Our approach consists in two steps. First, the properties, in terms of size and excitation energy, of the  
remnants of the collision are compared as obtained in three different abrasion models. 
In a second step, the decay of the remnants is computed within a dynamical model based on the Langevin approach. Various de-excitation channels 
are open to the decay of the highly-excited systems produced in the first stage of the collision, going from light-particle 
evaporation, intermediate-mass fragment (IMF) emission, fission, multifragmentation, and up to vaporisation. In the present model, the 
Langevin code is restricted to the spectator decay by evaporation and
fission. Other channels are not treated here. The dynamical 
results are compared to the predictions by the abrasion-ablation statistical model ABRABLA \cite{gaimard:1991,kelic:2008} which has 
shown successful in predicting the spectator decay in the beam energy range from about 100 to several thousands of MeV/nucleon.  
As a test case for our new dynamical framework, we study the reaction $^{208}$Pb+$^{208}$Pb at 158 GeV/nucleon energy 
($\sqrt{{s}_{NN}} =$ 17.3 GeV) measured at the CERN SPS at various
centralities \cite{schlagheck:2000}, and for which electromagnetic distortions were observed in 
the kinematical pion distributions.

%%% A. Kelic, M.V. Ricciardi and K.-H. Schmidt, arXiv_nucl-th/0906.4193v1; 
%%% H. Schlagheck (WA98 Collaboration), Nucl. Phys. A 663, 725 (2000).
%
%The "abrasion-ablation" is rather an approach than a definite well
%fixed model. So details differ from paper to paper.

%%%%%%%%%%%%%%%%%%%%%%%%%%%%%%%%%%%%%%%%%%%%%%%%%%%%%%%%%%%%%%%%%%%%%%%%%%%%%%%%%%%%%%%%%%%%%%%%%%
%%%%%%%%%%%%%%%%%%%%%%%%%%%%%%%%%%%%%%%%%%%%%%%%%%%%%%%%%%%%%%%%%%%%%%%%%%%%%%
%%%%%%%%%%%%%%%%%%%%%%%%%%%%%%%%%%%%%%%%%%%%%%%%%%%%%%%%%%%%%%%%%%%%%%%%%%%%%%
\section{Abrasion models}\label{abramod}
%%%%%%%%%%%%%%%%%%%%%%%%%%%%%%%%%%%%%%%%%%%%%%%%%%%%%%%%%%%%%%%%%%%%%%%%%%%%%%
%%%%%%%%%%%%%%%%%%%%%%%%%%%%%%%%%%%%%%%%%%%%%%%%%%%%%%%%%%%%%%%%%%%%%%%%%%%%%%
%%%%%%%%%%%%%%%%%%%%%%%%%%%%%%%%%%%%%%%%%%%%%%%%%%%%%%%%%%%%%%%%%%%%%%%%%%%%%%
%
The first stage of the collision between two heavy ions at ultra-relativistic energies 
($\sqrt{s_{NN}}\ge 5~$GeV) is very fast and energetic. As mentioned above, so 
far not much attention was paid to the description of the spectator decay. According to the
suggested importance of the time evolution of the latter \cite{rybicki:2007} on pion trajectories, 
we focus on this aspect.
Since the system considered in the study ($^{208}$Pb+$^{208}$Pb) is symmetric, the de-excitation of one 
remnant (from either the projectile, or the target), only, has 
to be explicitly computed. The calculation can, of course, easily be generalized to asymmetric entrance channels.\\
Three different abrasion models are proposed including, with increasing sophistication: a 
purely geometrical and macroscopic picture based on the Liquid Drop Model (LDM), the abrasion model ABRA of Gaimard  and Schmidt
\cite{gaimard:1991}, and the microscopic theory of Glauber \cite{hufner:1975}. The masses and excitation energies 
as predicted for the heavy remnant (hereafter prefragment or spectator) as a function of impact parameter in these three models are first compared. Next, they are used as the input 
for the calculation of its decay.\\
We note that the abrasion picture is valid for beam velocities larger
than the Fermi velocity. 
Its 
upper limit of applicability was never tested to our knowledge. Thus, the present work is also an interesting exploratory investigation 
in this respect.

%%%%%%%%%%%%%%%%%%%%%%%%%%%%%%%%%%%%%%%%%%%%%%%%%%%%%%%%%%%%%%%%%%%%%%%%%%%%%%
\subsection{Geometrical macroscopic approach} \label{abrageo}
%%%%%%%%%%%%%%%%%%%%%%%%%%%%%%%%%%%%%%%%%%%%%%%%%%%%%%%%%%%%%%%%%%%%%%%%%%%%%%

Right after the collision the spectator prefragment can experience a very exotic shape, which relaxes quickly towards a spherical 
configuration. In the here-proposed simplest approach, the excitation energy of the prefragment is equal to the Liquid Drop 
deformation energy \cite{bowman:1973}, calculated as the difference between the liquid drop energy of the deformed and spherical shapes.
\begin{eqnarray}\label{eqexc}
    E^*=&&E_{LDM}({\rm deformed \,\,\, spectator})\nonumber\\
    &&-E_{LDM}({\rm spherical \,\,\, spectator}),
\end{eqnarray}

The crucial point is the calculation of the surface energy of the
possibly exotic prefragment shapes as a function of impact parameter $b$, 
defined as the distance between the centers of the colliding nuclei. Depending on $b$, the way the aforementioned cut-off takes place may 
be different. Three scenarios are considered for determining the shape of the remnant of the collision. Its volume (equivalently, mass) is 
directly related to the number of nucleon removed or 'abraded', assuming a constant nuclear matter density. The Unchanged Charge Density 
(UCD) assumption is further used to determine its 
neutron and proton numbers according to: $A_{initial}/Z_{initial} = A_{spectator}/Z_{spectator}$, where the subscript {\it initial} refers to the 
projectile (equivalently, target).

Let us consider the simple geometrical situation of a sphere cut off by a plane (hereafter 'sphere-plane'). Then the final shape resembles a sphere 
without spherical dome. The corresponding form can be described as follows:
\begin{eqnarray}
 S(sphere) =&&\big \{ (\rho,z,\phi): 0\leq z\leq R, \nonumber\\
 &&0\leq\rho\leq\sqrt{R^2-z^2}, 0\leq \phi\leq 2\pi \big\}
\; .
\end{eqnarray}
The spherical cap:
\begin{eqnarray}
 W=&&\big \{ (\rho,z,\phi): b\leq z\leq R, \nonumber\\
 &&0\leq\rho\leq\sqrt{R^2-(b+z)^2}, 0\leq \phi\leq 2\pi\big\}
\; .
\end{eqnarray}
The deformed spectator:
\begin{equation}
 \Omega=S(sphere)/ W ; .
\end{equation}
The volume of the spectator in 'sphere-plane' scenario:
\begin{eqnarray}
V_{\Omega} &=& \int\int_{ S(sphere)/ W}\int d V(\rho,z,\phi)\nonumber\\
&=& \int_{0}^{2\pi}d\phi\int_{0}^{b}dz\int_{0}^{\sqrt{R^2-z^2}}d \rho
\end{eqnarray}
and the surface:
\begin{equation}
 S_{\Omega}(b)=2\pi \int_{0}^{b} dz \sqrt{R^2-z^2} \; .
\end{equation}
The volume of the spherical cap:
\begin{eqnarray}
 V_W&=&\int\int_{W}\int d V(\rho,z,\phi)\nonumber\\
 &=& \int_{0}^{2\pi}d\phi\int_{b}^{R}dz\int_{0}^{\sqrt{R^2-(b+z)^2}}
 \rho d\rho,
\end{eqnarray}
where $S$ is the surface of the sphere with radius $R$ and $dS$ is a
two-dimensional differential element of the sphere volume.
The surface of the sphere described by the function: $x^2+y^2+z^2=R^2$ cut
by the plane $z=b$ is calculated as:
\begin{eqnarray}
 &&S(def)
 =
 \int\int_D \sqrt{1+\bigg(\frac{\partial z}{\partial x}\bigg)^2+\bigg(\frac{\partial z}{\partial y}\bigg)^2}dxdy \nonumber\\
 &&=
 \int_{-\sqrt{R^2-b^2}}^{\sqrt{R^2-b^2}}\int_{-\sqrt{R^2-b^2-x^2}}^{\sqrt{R^2-b^2-x^2}}
 \frac{R}{\sqrt{R^2-x^2-y^2}} dydx, \nonumber\\
\end{eqnarray}
where the impact parameter $b$ spans the range $(0,R)$ and $D$
is a projection of the sphere on the OXY plane.
\begin{figure}[!bt]
   \begin{center}
     \includegraphics[scale=0.60]{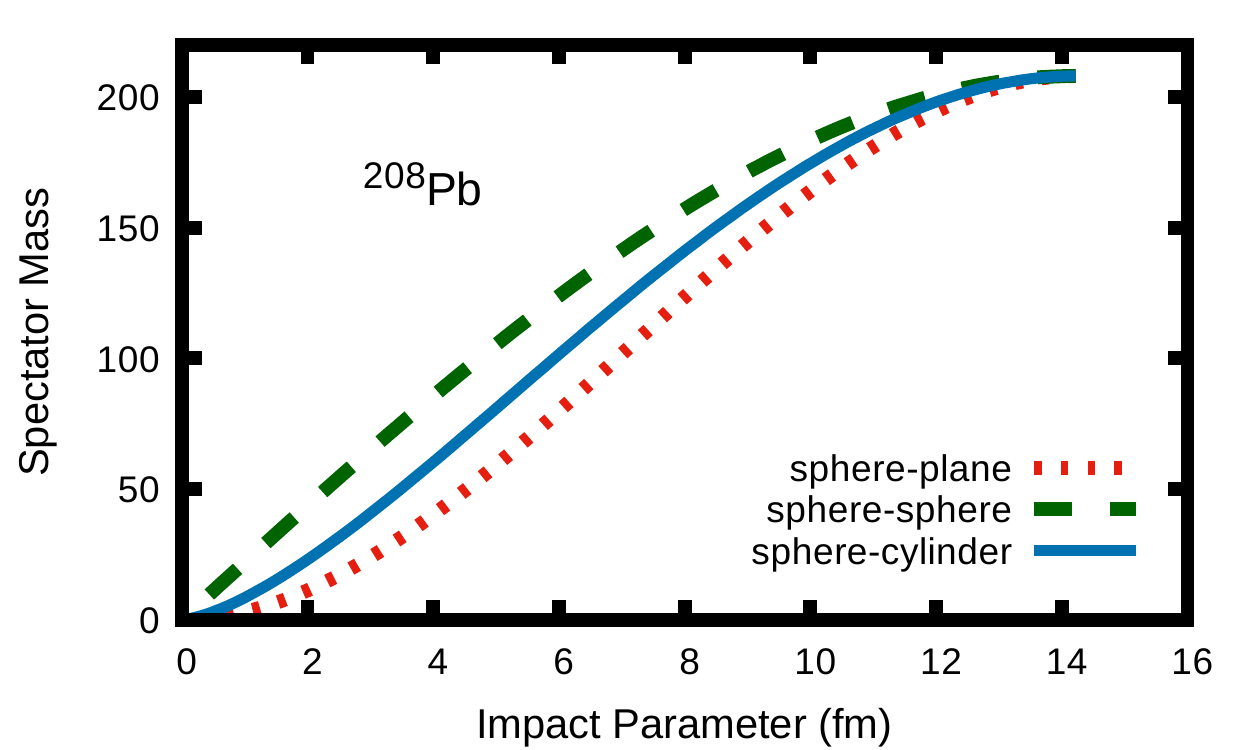}
     \caption{(Color on-line) Mass of the spectator produced in 
     three geometrical scenarios described in the text as a function of the impact parameter for $^{208}$Pb+$^{208}$Pb collisions.
             }
                                                                 \label{fig.01}
   \end{center}
\end{figure}

The second geometrical situation we wish to consider corresponds to two spheres that penetrate each other (hereafter 'sphere-sphere'). 
Hence, the final shape looks as a sphere without a double spherical dome like {\it e.g.} a lens. Assuming that the two nuclei 
are identical, the surface of the spectator prefragment will be similar to that of the initial sphere, but its volume will 
be different.

The third geometrical situation considered in this work takes, in an effective way, into account the dynamics of the 
process. It is based on the idea that the collision is very fast, and the nuclei 
interact with each other as a finite-size bullet grazing a sphere, or said differently, the projectile scraps the 
target with a cylinder (hereafter 'sphere-cylinder'). The shape obtained under this assumption is rather exotic, complex to calculate
geometrically; though it seems to be most realistic for the beam energies typical at the CERN SPS \cite{westfall:1976}.

The equations governing the calculation of the aforementioned three geometrical configurations, and namely the surface, 
are detailed in Appendix \ref{a1}. Figure~\ref{fig.01} presents the dependence of the mass of the spectator prefragment
on the impact parameter in the $^{208}$Pb+$^{208}$Pb collisions for the three scenarios. Since, for a given impact parameter, the corresponding 
shapes have distinct volumes, the remnant mass obtained in the three cases is different, as well as its surface, neutron and proton numbers.

We note that, under the assumption that the nucleus equilibrates its shape from deformed to spherical quickly compared to the time 
of its decay, we neglect the explicit treatment of the dynamics of this first shape relaxation process. That is, for each impact 
parameter, the resulting prefragment mass, charge and excitation energy are calculated based on geometrical considerations only, 
as detailed below.

Once the shape of the spectator prefragment is established for a given scenario and impact parameter, a purely macroscopic picture 
is proposed in this work in order to determine its excitation energy. It is assumed that a sound estimate of the latter can be obtained 
from the deformation energy as predicted by the LDM within the sudden cut-off approximation. The Lublin-Strasbourg Drop (LSD) 
model \cite{pomorski:2003,dudek:2004} is used in this study. The main contribution to the deformation energy of a nucleus 
is given by the surface energy; the Coulomb and curvature energies giving second-order corrections. In this work, we therefore approximate the 
prefragment deformation energy with its surface energy. For a deformed nucleus with mass $A$ and charge $Z$, the LDM 
surface energy reads:
\begin{equation}
      E_{\rm surf.}(A,Z;def)
      =
      b_{\rm surf.}\,(1 - \kappa_{\rm surf.} I^2\,)\,A^{2/3} 
      B_{\rm surf.}(def) .
                                                                \label{eqn.05} 
\end{equation}
where $I=(A-2Z)/A$. The deformation-dependent term is defined as the surface energy of the 
deformed body normalized to that of a sphere of the same volume:
\begin{equation}\label{eqn.05a}
 B_{\rm surf.}(def) = \frac{S(def)}{S(sphere)}.
\end{equation}
Details about the LDM formulas, and its LSD implementation and parameters can be found in Appendix~\ref{a2}.\\

As introduced in this section, in the proposed simple geometrical macroscopic 
abrasion model, the excitation energy would be given by Eq.~\ref{eqexc} where $E_{LDM}$ is approximated 
by $E_{surf}$ in the LSD parameterization. According to \cite{oliveira:1979} the excitation energy derived from 
the surface-energy excess of the deformed prefragment is too low. Guided by the results of \cite{schmidt:1993}, 
in the present work, the excitation energy considered in the framework of the purely geometrical macroscopic 
approach is taken as {\it two times} the value calculated with Eq.~\ref{eqexc}.
%L.F. Oliveira et al., PRC 19 (1979) 826

\begin{figure}[!bt]
   \begin{center}
     \includegraphics[scale=0.56]{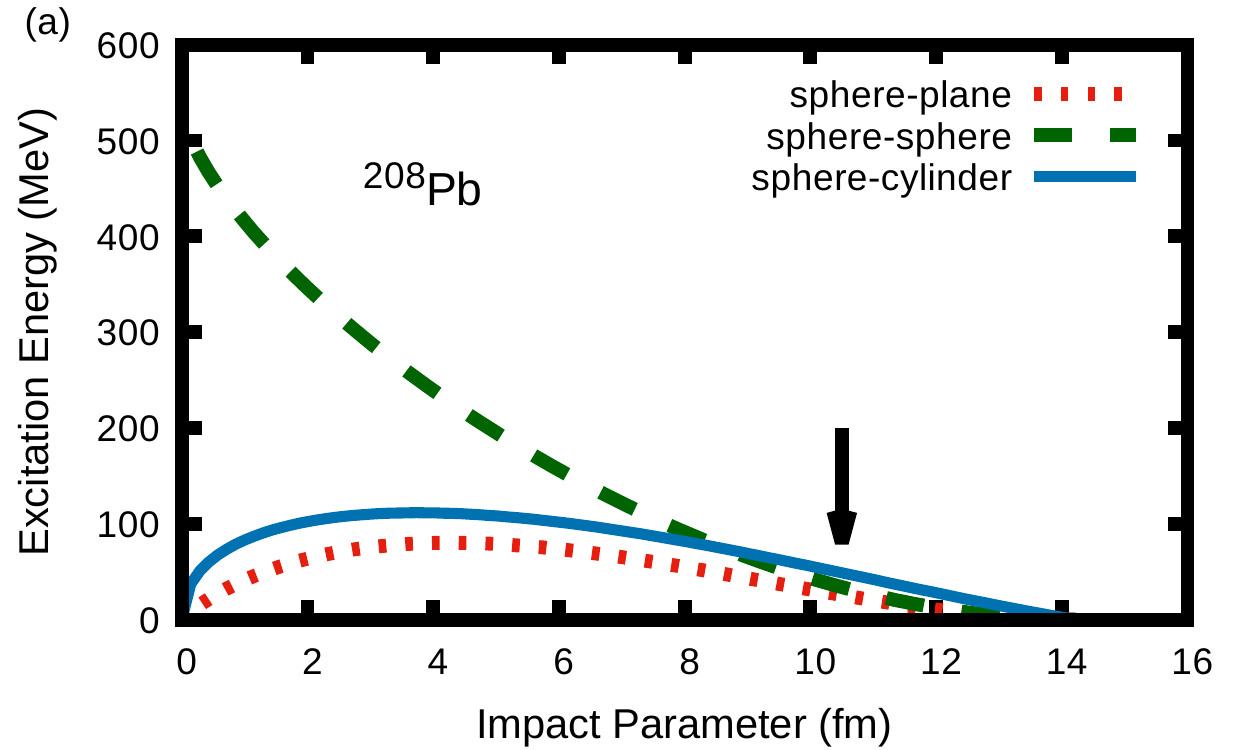}
     \includegraphics[scale=0.56]{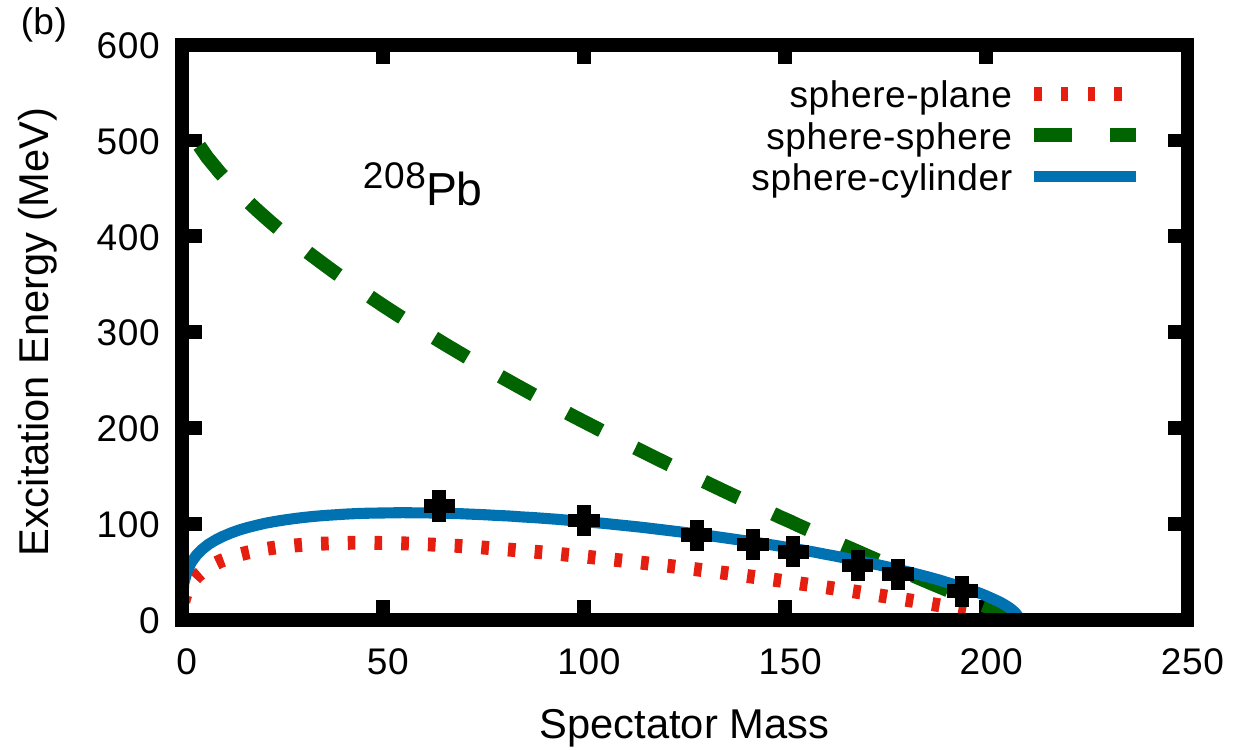}
    \caption{(Color on-line) Dependence of the prefragment excitation energy for the three geometrical 
    scenarios described in the text, on impact parameter (a), and on mass (b). 
     The arrow indicates the impact parameter considered in 
     \cite{rybicki:2007}. Crosses in panel (b) mark the nuclei selected as typical examples for further dynamical calculations.
             }
                                                                 \label{fig.02}
   \end{center}
\end{figure}

Figure~\ref{fig.02} shows the excitation energy predicted by the macroscopic approach for the three considered geometrical 
scenarios, as a function of 
impact parameter (a), and spectator mass (b). It is observed that, depending on the geometrical abrasion hypothesis, the  
excitation energy can have a rather different behavior, and take substantially different 
values\footnote{Preliminary results reported in \cite{mazurek:2017a}
  suffered from some technical issue, yielding somewhat erroneous 
numerical values. Yet, the issue, solved here, did not affect the main outcome and conclusion of that work.}. The largest 
excitation energy is predicted for close-to-central collisions in the 'sphere-sphere' picture, since the spectator object 
after the collision has the most curved and deformed shape. The excitation energy expected within this model can reach up to 
500~MeV, where multifragmentation-like processes are very likely to contribute. The 'sphere-plane' and 'sphere-cylinder' 
scenarios predict excitation energies below about 100~MeV and 150~MeV, respectively, for semi-central collisions. This $E^*$ regime 
is within the domain of applicability of the stochastic Langevin approach restricted to the competition between evaporation and fission.\\

The correlation between prefragment mass and excitation energy is crucial in the calculation of its time evolution and decay. Note that, within the
here-proposed simplest geometrical macroscopic picture, the correspondence between excitation energy and impact parameter (equivalently, mass) 
is a {\it one-to-one} correspondence. Furthermore, the angular momentum of the spectator prefragment is assumed to 
be negligible; this approximation is reasonable for the present exploratory study, and can be easily leveled-off in future. 
The black crosses in Fig.~\ref{fig.02} mark the prefragments which were selected for further investigation 
of the geometrical macroscopic picture and combined to the dynamical Langevin approach. The 
'sphere-cylinder' scenario is chosen for this investigation, as it seems the most realistic assumption. 
The arrow at $b \approx$~10~fm 
in Fig.~\ref{fig.02} a) indicates the impact parameter considered in \cite{rybicki:2007} for studying the influence of the spectator 
electromagnetic field on pions.

\begin{figure}[!bt]
   \begin{center}
     \includegraphics[scale=0.60]{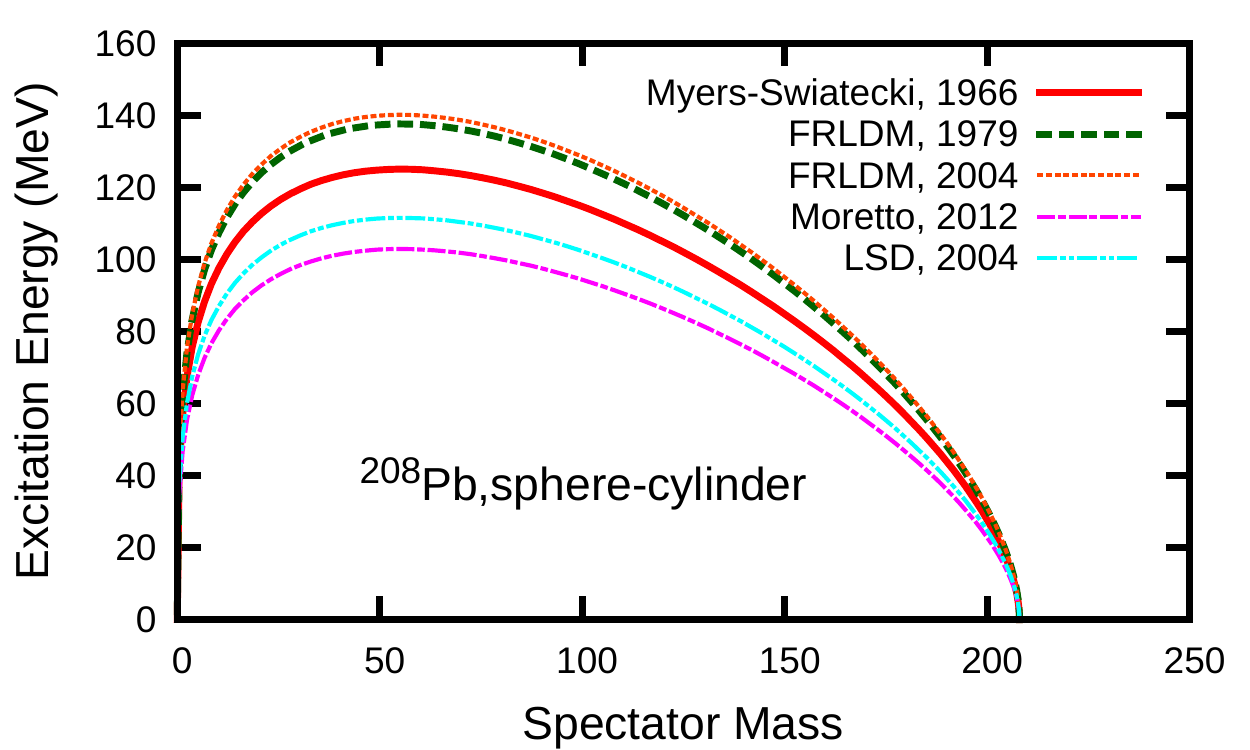}
    \caption{(Color on-line) Dependence of the spectator prefragment excitation energy on its mass for various LDM realizations: 
    Myers-Swiatecki, 1966 \cite{myers:1967}, FRLDM, 1979 \cite{krappe:1979}, FRLDM, 2004 \cite{moller:2004}, 
    Moretto, 2012 \cite{moretto:2012}, and LSD, 2004 \cite{pomorski:2003}.
             }
                                                                 \label{fig.03}
   \end{center}
\end{figure}

We note that the predicted value of the prefragment excitation energy can also be influenced by the 
specific LDM parameterization used. This is illustrated in Fig.~\ref{fig.03} which displays the correspondence between 
prefragment mass and excitation energy for the 'sphere-cylinder' scenario, as obtained with most popular implementations 
of the LDM. The largest excitation energies are predicted by the Finite Range Liquid Drop Model with
the latest set of parameters (FRLDM, 2004 \cite{moller:2004}), while the lowest excitation energies are obtained 
with Moretto's prescription (Moretto, 2012 \cite{moretto:2012}). The LSD model used in this work predicts values 
close to the lower boundary, hardly exceeding 100~MeV. The spread in excitation energy depending on the LDM used can reach 40~MeV. That shall give an idea about the uncertainty range of the spectator excitation energy predicted within the geometrical macroscopic 
abrasion model proposed in this work.

%----------------------------------------------------------------------------------
\subsection{Statistical abrasion model of Gaimard-Schmidt} 
%----------------------------------------------------------------------------------

The geometrical picture sketched above is also the
basis of the abrasion model (ABRA) of Gaimard-Schmidt \cite{gaimard:1991} widely used in the field.  
For a given abraded mass, the protons and neutrons are assumed to be removed randomly from the 
projectile, and statistical fluctuations given by the hyper-geometrical distribution yield the neutron-to-proton 
ratio of the prefragment spectator.

A major difference as compared to the picture outlined in the previous section is that the calculation of the prefragment excitation energy implemented in the ABRA code is not 
based on a geometrical macroscopic approach. Rather, it is given by 
the energy of the vacancies created in the single-particle (s.p.) levels with respect to the Fermi surface \cite{schmidt:1993}. 
In this sense, ABRA accounts for microscopic effects in the entrance channel of the reaction, contrary to the macroscopic 
picture of the previous section. The excitation energy computed this way
leads to higher values than 
those based on the 
surface energy excess \cite{gaimard:1991}. Though, comparison with experiment suggests that the calculation based on s.p. levels 
vacancies still gives a too low excitation energy. The result was thus further empirically adjusted by 
multiplying the theoretical value by a factor of two \cite{schmidt:1993}. 
This is what is implemented in ABRA. The deviation 
between the theoretical calculation and the empirically adjusted value may
be due to friction effects or final-state 
interactions \cite{schmidt:1993, oliveira:1979, gaimard:1991}. The
angular momentum of the pre-fragment in ABRA is calculated as the
sum of the angular momenta of the nucleons removed in the collision \cite{jong:1997}. 

To describe the entire reaction, from the early collision up to the final cold products are reached, the ABRA code is 
usually combined with the statistical evaporation model ABLA \cite{kelic:2008}. 
In its most general form, the ABRABLA code consists of three stages: (1) abrasion (ABRA), (2) if the 
temperature of the remnant after abrasion is above a limiting value (around 4.5~MeV), the system breaks up in several more or less 
heavy intermediate products \cite{schmidt:2002}, (3) de-excitation (ABLA) of the heavy remnants from stage (1) or (1)+(2). We 
note that in the first stage of the reaction, the code considers that abrasion can be induced by, either nuclear or electromagnetic, 
interactions. The latter are confined to large impact parameters, and their probability increases with the charge of the colliding 
ions. They involve small, below about 30~MeV, excitation energy. We finally emphasize that ABRABLA computes the decay of the heavy-ion 
remnant of the collision, only. The decay of the nuclear matter in the overlapping zone is not followed.

In the remainder of this work, unless explicitly specified, we consider only those events from ABRABLA which do not pass by stage 
(2); that is, the very-highly excited prefragments experiencing a break-up process prior 'standard' de-excitation are excluded. 
Events with IMF emission are not considered neither, as well as we do overlook electromagnetic-induced reactions. All in all, 
we restrict to those events which undergo most 'standard' low-energy de-excitation process, leading to, either evaporation of light particles and formation of a heavy evaporation residue (ER), or fission possibly 
accompanied by light-particle emission. This restriction is chosen in order to permit most meaningful 
comparison with the Langevin calculations detailed later below, and which model the ER and fission channels, only. We emphasize 
that this selection excludes very central collisions. That is welcome also, since the geometrical abrasion picture outlined 
above makes certainly most sense for the more peripheral collisions. Finally, the study of Ref.~\cite{rybicki:2007} about the 
influence of spectator-induced electromagnetic fields on pion trajectories was done for $b \approx$~10.5 fm, belonging to the peripheral
collision domain.\\        

\begin{figure}[!bt]
   \begin{center}
     \includegraphics[scale=0.60]{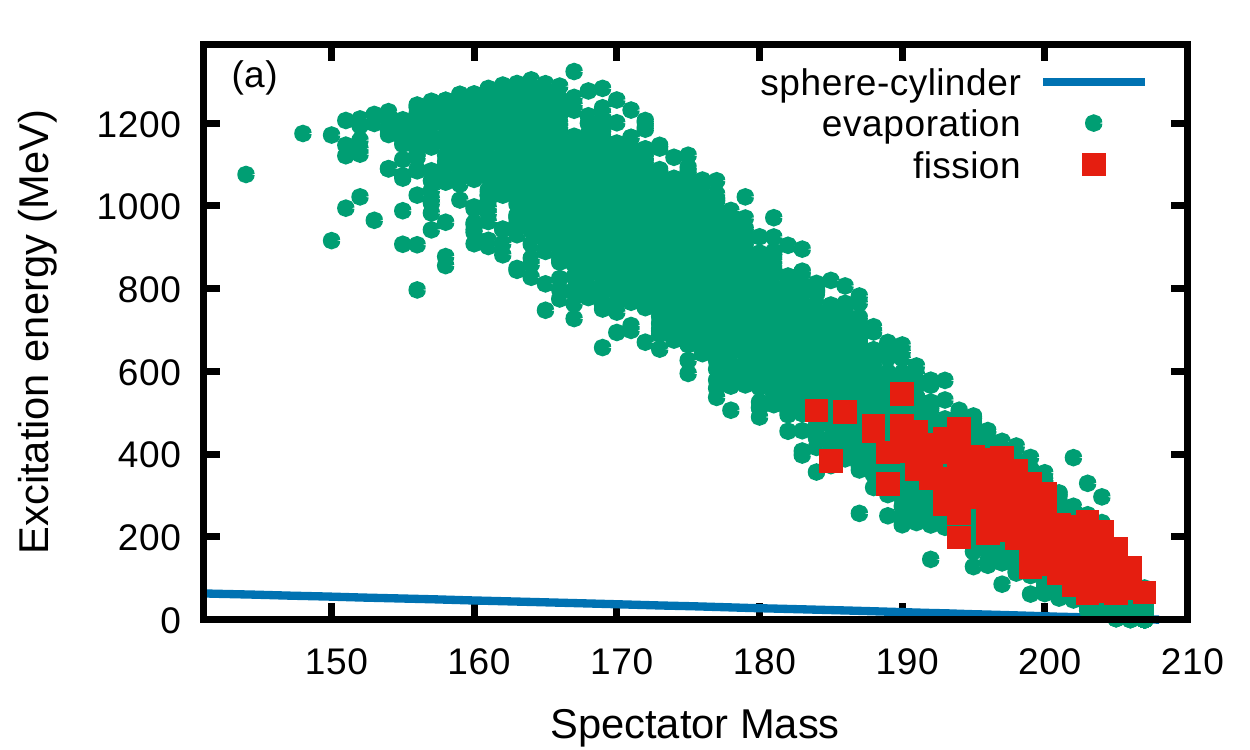}
     \includegraphics[scale=0.60]{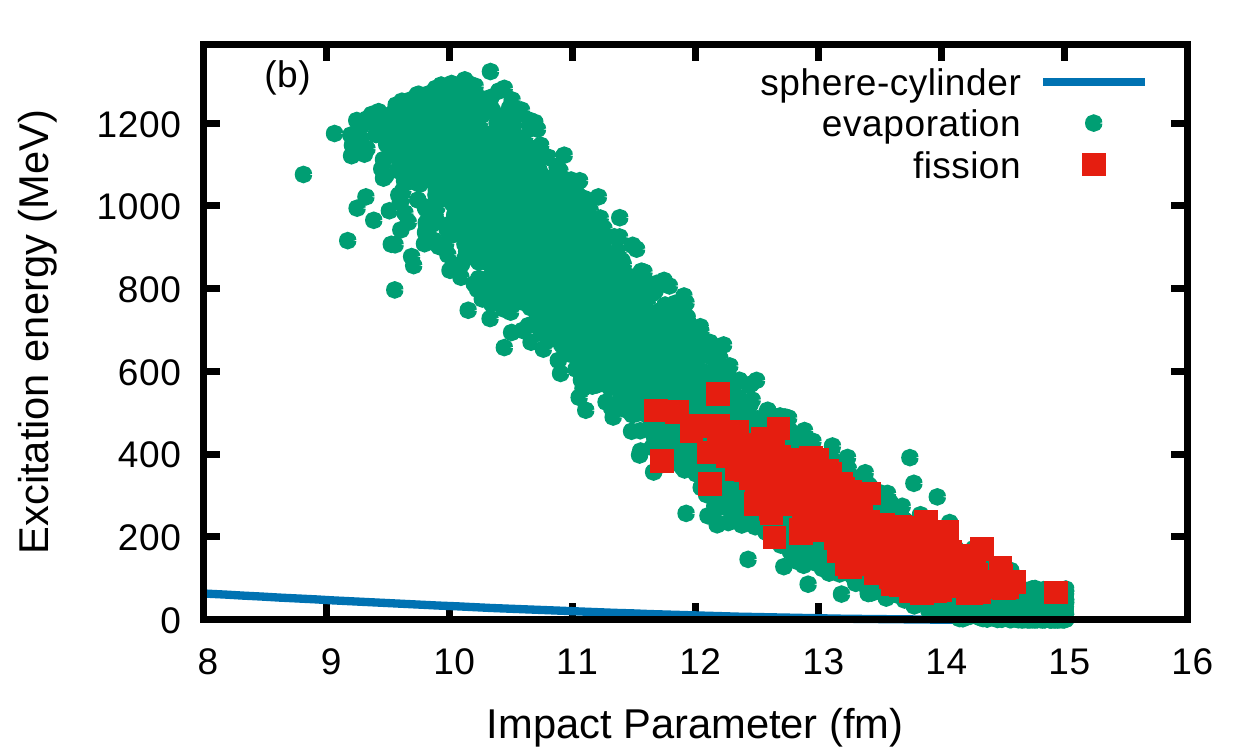}
     \caption{(Color on-line) Correlation between the prefragment excitation energy on its mass (a), and on 
     impact parameter (b) for events ending with the formation of a single heavy ER, or with fission, as 
     predicted by the ABRABLA code. For comparison, the correlation compted in the geometrical macroscopic 
     model for the 'sphere-cylinder' is shown again.}
                                                                 \label{fig.04}
   \end{center}
\end{figure}

Figure~\ref{fig.04} shows the correlation between the prefragment excitation energy and mass (a), and the prefragment excitation energy 
and impact parameter (b) as predicted with ABRABLA. Events ending in fission or the survival of a single heavy evaporation residue are shown separately. The full line displays again the dependence of the excitation energy on mass and impact parameter, 
respectively, as determined with the geometrical 'sphere-cylinder' picture in the previous section. It is clear that ABRA 
gives much higher excitation energies than the geometrical macroscopic calculation, as shown already in Ref.~\cite{gaimard:1991}. The geometrical
calculation extends down to the lowest prefragment masses (see also Fig.~\ref{fig.02}b), while such events are not present for ABRA as they usually imply passage through the break-up stage (2) which we disregard.

In the present work, in addition to analyzing the predictions of the
ABRABLA code as such, we also construct a 'hybrid' model, by 
using the results of ABRA as an input for the dynamical calculations
within the Langevin approach presented below. 
That will permit to investigate the influence of i) the modeling of 
the prefragment properties in the abrasion stage, and ii) the difference between the de-excitation path based on a statistical or dynamical model.

The ABRABLA code has shown successful over a wide beam energy range from about 100 to several thousands of MeV/nucleon. To our knowledge, 
it was never tested in the ultra-relativistic energy domain of this work. As mentioned above, the ABRA stage may implicitly imply some 
degree of friction between the colliding nuclei. Whether friction is still present at ultra-relativistic velocities is not obvious. Hence, 
any attempt to probe the upper energy limit for the validity of the
ideas behind ABRA is worthy consideration. 
The exploratory work done here suggests these ideas to be rather robust, as will be seen further below.

%-------------------------------------------------
\subsection{Microscopic Glauber abrasion model}
%-------------------------------------------------

At relativistic beam energy, the abrasion cross section is most often computed, like in the two previous sections, assuming 
a geometrical picture for the impact parameter distribution. A more elaborate 
prescription was proposed with the Glauber theory 
of multiple scattering \cite{hufner:1975}. The Glauber model is a microscopic approach which uses the 
matter densities calculated for protons and neutrons removed from the nucleus.
The nuclear matter densities are obtained from the Woods-Saxon potential, or 
any other s.p. potential. In the present work, the Hartree-Fock-Bogolyubov (HFB) method is used 
assuming a spherical shape for $^{208}$Pb \cite{dobaczewski:1996}.

The abrasion cross section from the Glauber model has the form:
\begin{equation}
\sigma_{abr}^{i}(N) = {A\choose N } 2 \pi
\int_{0}^{b_{max}} b db \left( 1 - P_{i}(b) \right)^N  
                        \left( P_{i}(b) \right)^{A_i - N} \; ,
\label{Glauber_formula}
\end{equation}
where $A_i$ is the mass number of one of the colliding nuclei (indexed $i$, $j$), $N$ is the number of 
abraded nucleons and
\begin{equation}
P_{i}(\vec{b}) =\int d^2 D_i(\vec{s}) 
\exp(-A_j \sigma_{NN} D_j(\vec{s}+\vec{b})) \,
\label{abrasion_probability}
\end{equation}
with $i,j=1,2$ and $i \ne j$.
The density function of the colliding nuclei is
\begin{equation}
D_i(\vec{s}) = \int_{-\infty}^{\infty} dz \rho_i(\vec{s},z).
\label{thickness_functions}
\end{equation}
We take $\sigma_{NN}$=40 mb for practical calculations, and the HFB neutron/proton densities are 
used to calculate $D(\vec{s})$ and $P(\vec{b})$.\\

\begin{figure}[!bt]
   \begin{center}
     \includegraphics[scale=0.70]{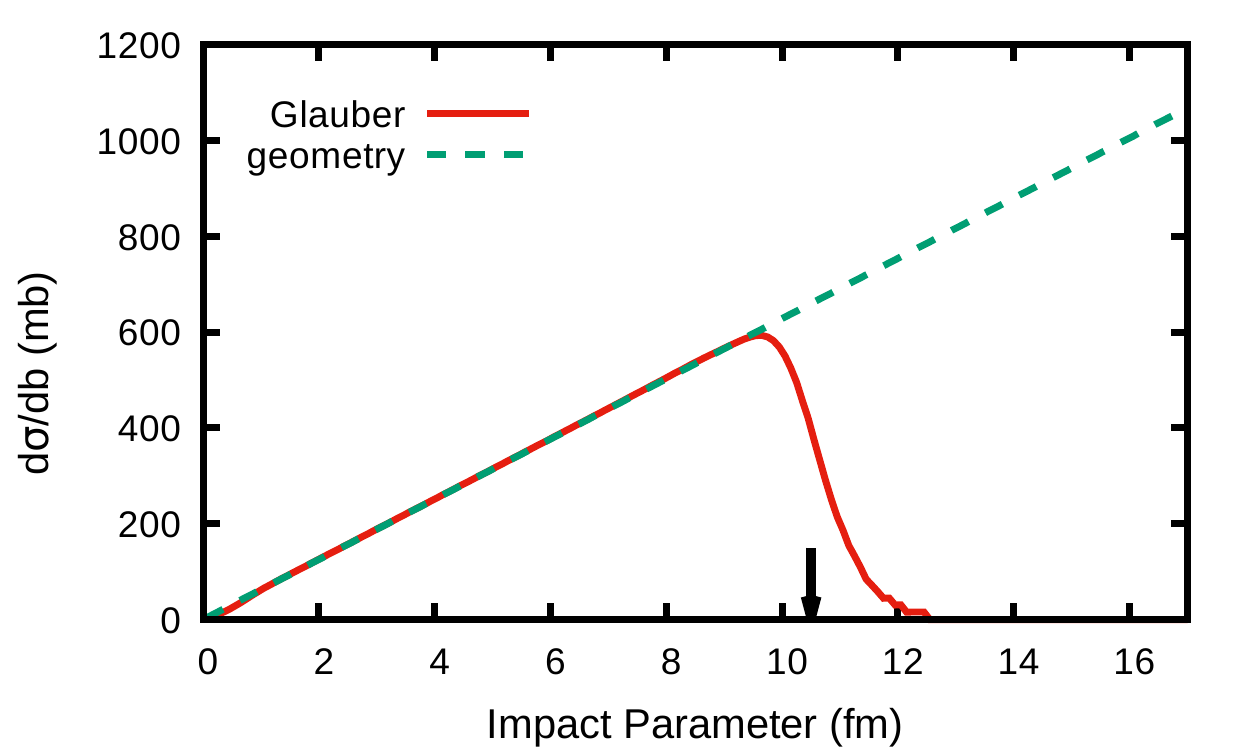}
      \caption{(Color on-line) Impact parameter distribution obtained from 
      the Glauber formula (Eq.~\ref{Glauber_formula})
      and from the geometrical prescription. The arrow marks the $b$ value studied in 
     \cite{rybicki:2007}.}
                                                                 \label{fig.04a}
   \end{center}
\end{figure}

In Fig.~\ref{fig.04a} the impact parameter distribution (equivalently, abrasion cross section) as 
predicted by the Glauber model is compared to the purely geometrical picture. Up to $b$=10~fm the Glauber distribution 
coincides with the latter, before it drops to zero at higher impact parameters. 

%{\bf A.S. has a problem with the abrasion cross section given by 
%Eq.(\ref{Glauber_formula})  when including large $N$.
%Also he gets strange dependence on $b$.
%Perhaps this should be extracted from ABRABLA for comparison.}

\begin{figure}[!bt]
%   \begin{center}
   \hspace{-0.9cm}
     \includegraphics[scale=3.25]{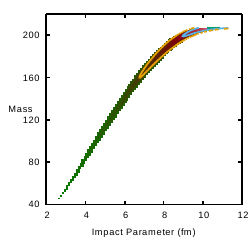}\hskip 3cm
      \caption{(Color on-line) Correlation between the spectator mass and impact parameter of the collision as obtained from 
      the Glauber formula (Eq.~\ref{Glauber_formula}).
             }
                                                                 \label{fig.04b}
%   \end{center}
\end{figure}

%In \cite{scheidenberger:2004} the Statistical Multifragmentation Model (SMM) 
%\cite{barz:1987,botvina:1995} was used. No time information is possible in
%the latter approach.

As for the prefragment excitation energy, in our implementation of the Glauber model, either
the Gaimard-Schmidt \cite{gaimard:1991,schmidt:1993}, or the Ericson \cite{ericson:1960} approach can be used. 
Depending on this choice, the mean excitation energy imparted to the remnant ranges from 10 to 20~MeV
per abraded nucleon. The excitation energy in our Glauber model is then taken as:
\begin{equation}
E_{exc} =  N \Delta E   \; ,
\label{excitation_energy}
\end{equation} 
where $N$ is the number of abraded nucleons, which makes sense provided that $N \ll A$, {\it i.e.} the number of holes
is small, as it is the case in peripheral collisions. For practical calculations
 we consider $\Delta E =$ 10~MeV. The correlation between the prefragment excitation energy and its mass is displayed in 
 Fig.~\ref{fig.04b}. When compared to the geometrical (Fig.~\ref{fig.01}) and ABRA (Fig.~\ref{fig.04}) results, our implementation of the Glauber approach yields a
 less steep decrease of prefragment mass with decreasing impact
 parameter, and leads to lower excitation energies.

\section{Dynamical decay of the spectator prefragment}  
%%%%%%%%%%%%%%%%%%%%%%%%%%%%%%%%%%%%%%%%%%%%%%%%%%%%%%%%%%%%%%%%%%%%%%%%%%%%%%
%%%%%%%%%%%%%%%%%%%%%%%%%%%%%%%%%%%%%%%%%%%%%%%%%%%%%%%%%%%%%%%%%%%%%%%%%%%%%%
%
%%%%%%%%%%%%%%%%%%%%%%%%%%%%%%%%%%%%%%%%%%%%%%%%%%%%%%%%%%%%%%%%%%%%%%%%%%%%%

To describe the decay of the hot remnant formed in the abrasion stage, a
statistical model is most commonly used. 
In the present work, we propose to innovatively extended the description
of the second (ablation) stage with a dynamical model.

The time evolution of the fissioning nucleus is described within the 
stochastic approach \cite{kramers:1940,abe:1996,frobrich:1998}. Most relevant degrees of freedom are 
introduced as collective coordinates, and their evolution with time is treated as the motion of Brownian particles, which interact stochastically with the larger number of internal degrees of freedom constituting a surrounding 'heat bath'. The details of the approach can be found in \cite{mazurek:2017}, and references therein; only the main features are given below.

In the present implementation of the stochastic method, four collective coordinates are considered. Three out 
of them define the shape of the nucleus, while the fourth defines its orientation in space. The coordinates 
${\rm \bf q}=(q_1,q_2,q_3)$ are connected to, respectively, elongation, neck thickness and left-right asymmetry 
\cite{adeev:2005, nadtochy:2005}. They are based on the popular Funny-Hills $(c,h,\alpha)$ nuclear-shape parameterization 
\cite{brack:1972}. The collective coordinate $q_4$=$K$ is taken as the projection of the angular momentum $L$ of the nucleus 
onto the fission axis, varying in the range (-$L$,+$L$). In the present work, whenever the 
dynamical stage is combined with the geometrical macroscopic or the Glauber abrasion models,   
the angular momentum imparted to the heavy remnant is not evaluated, and $L$ is set to zero in first approximation. The 
dynamical calculation is hence three-dimensional, only. On the contrary, when the dynamical stage is fed with the input 
from the ABRA abrasion model, the angular momentum is directly taken from ABRA. The second stage is then 
effectively a four-dimensional calculation. \footnote{The difference in angular-momentum treatment depending on the 
abrasion model has no significant influence on the observables of our study, as the angular 
momentum imparted to the heavy prefragment remains limited to a few $\hbar$ on average \cite{gaimard:1991}.}
Within the present stochastic approach, the time evolution of the shape coordinates is given by the solution of the 
Langevin equation:
\begin{eqnarray}\label{eqlan} 
\frac{dq_i}{dt}
&=&
\sum_{j}^{} \mu_{ij}(\vec{q})p_j, \nonumber\\
\frac{dp_i}{dt}
&=&
-\frac{1}{2} \sum^{ }_{j,k} \frac{d\mu_{ij}(\vec{q})}{dq_i} p_jp_k
-
\frac{d F(\vec{q})}{dq_i}\nonumber\\
&-&
\sum^{ }_{j,k} \gamma_{ij} (\vec{q})\mu_{ij}(\vec{q})p_k
+
\sum^{ }_{j}\theta_{ij}(\vec{q})\xi_j(t)\label{eqp},
\end{eqnarray}
\noindent
for ${\rm \bf q}$ the vector of collective coordinates, 
and ${\rm \bf p}$ the vector of conjugate momenta. 
The evolution is governed by driving potential, friction and inertia
forces, all explicit functions of deformation. 
A similar equation, with some variation to take the specificity of this
model into account, holds for $q_4$=$K$ \cite{mazurek:2017}.

\begin{figure*}[!hbt]
   \begin{center}
     \includegraphics[scale=0.550]{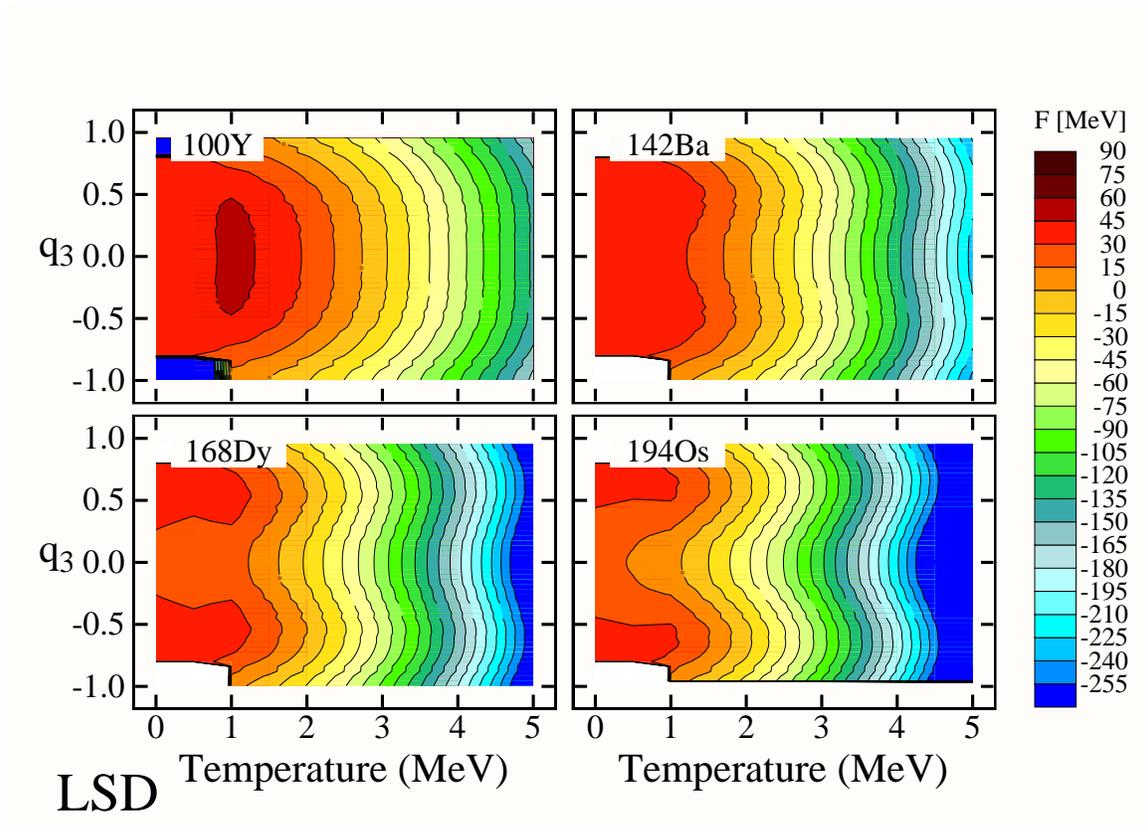}
     \caption{Fission barrier in the ($T$, $q_3$) plane for a sample of
       nuclei. Here the barrier is defined as the difference between the
       free energy at the conditional ($q_3$-constrained) 
saddle point and the ground-state equilibrium point. The landscapes are presented for $^{100}$Y, $^{142}$Ba, $^{168}$Dy and $^{194}$Os.}
                                                                 \label{fig.05}
   \end{center}
\end{figure*}

The driving potential is given by the $\rm F({\bf q})=V({\bf q}) - {\it a}({\bf q}) T^2$ Helmholtz free energy, 
where the bare potential energy $\rm V({\bf q})$ is obtained from the LSD model, and $T$ is the temperature of 
the decaying system. The inertia tensor $m_{ij}({\bf q})$ is calculated under the Werner-Wheeler approximation of 
an incompressible irrotational flow \cite{davies:1976}. The friction tensor $\gamma_{ij}({\bf q})$ is derived from the 
one-body {\it wall-plus-window} prescription \cite{blocki:1978} modified in order to account properly for chaocity 
of the nucleon movement  inside the deformed nucleus \cite{chaudhuri:2001}. Fluctuations 
are modeled by the random force  $\theta_{ij}$ related to friction by the Einstein relation 
$\rm \sum \theta_{ik}\theta_{kj} = T \gamma_{ij}$. The temperature of the system $ T$ is determined by 
the Fermi-gas model formula 
\begin{equation}\label{temp}
 \rm T=(E_{int}/{\it a})^{1/2}, 
\end{equation}
where $\rm E_{int}$ is the internal excitation energy of the nucleus, and $a$ is the level-density parameter. 

Along its dynamical evolution the system may evaporate particles and $\gamma$-rays. The Master equation governing 
this process is coupled to the multi-dimensional Langevin equation \cite{mavlitov:1992}. Both set of equations are 
solved together within a Monte Carlo framework \cite{adeev:2005} time-step by time-step ($\Delta t = 10^{-23}$s). At 
each step, the properties of the decaying system (in mass, charge, excitation energy, and angular momentum wherever applicable) are recalculated 
taking into account its possible change of shape, or evaporation of a particle. If the nucleus, along its trajectory, 
is driven to a very elongated and necked-in shape, it splits into two fragments, {\it i.e.} fission occurs. Any particle emitted 
before this so-called scission point is then denominated 'pre-scission' particle. On the contrary, if the nucleus exhausts 
its excitation energy before reaching a scission shape, it ends in the state of a heavy evaporation residue (ER). 

Figure~\ref{fig.05} shows the fission barrier of some specific heavy
nuclei as a function of their left-right 
shape asymmetry and their temperature. The $T$-dependent barrier is defined as the difference between the free 
energy at the saddle point and at the equilibrium ground state. It represents the energy the nucleus has to 
possess in order to have a chance (classically) to fission. This threshold energy clearly depends on the how 
the nucleus splits {\i.e.} in fragments of either equal or different size. For example, the topography of the maps in 
Fig.~\ref{fig.05} suggests that $^{100}$Y will most likely fission into asymmetric fragments independent of the 
temperature (as the barrier to overcome is lower for $q_3$ non zero), while the heavier systems of the figure 
will most often experience symmetric fission. We note also that increasing temperature usually induces broader 
fragment  mass (equivalently, charge) distributions, as suggested by the softer driving potential landscapes 
in the $q_3$ asymmetry direction when $T$ is larger, as well as due to larger fluctuations.\\

The dynamical approach is aimed to model the second stage of the collision. In the present work, it is combined 
to the three abrasion models of Section~\ref{abramod}, leading to three different reaction softwares, called 
LSD-Lang, Glauber-Lang, and ABRA-Lang, respectively. For sake of comparison, calculations with the ABRABLA 
code will be presented as well, in which the ABRA abrasion model is combined to its companion decayl model ABLA. The latter has shown very powerful in describing the competition between the various open decay channels (evaporation and 
fission for the concern of this work), as well as the properties (mass, charge, energy) of the light-particle 
and heavy residue or fission fragment products. As compared to other statistical models, ABLA possesses some 
specific assets. These comprise a parameterization of the fission-decay width which accounts to some extend and 
in an effective way for friction effects along the path to fission, and the explicitly account of an 
elaborate empirical potential landscape for the determination of fission fragment properties \cite{kelic:2008}. Though, like any statistical model, the ABLA software does not give a direct access to a true time for the decay.

%
%%%%%%%%%%%%%%%%%%%%%%%%%%%%%%%%%%%%%%%%%%%%%%%%%%%%%%%%%%%%%%%%%%%%%%%%%%%%%%
%%%%%%%%%%%%%%%%%%%%%%%%%%%%%%%%%%%%%%%%%%%%%%%%%%%%%%%%%%%%%%%%%%%%%%%%%%%%%%
%%%%%%%%%%%%%%%%%%%%%%%%%%%%%%%%%%%%%%%%%%%%%%%%%%%%%%%%%%%%%%%%%%%%%%%%%%%%%%
\section{Results}

\subsection{Geometrical macroscopic abrasion coupled to dynamical decay}
%%%%%%%%%%%%%%%%%%%%%%%%%%%%%%%%%%%%%%%%%%%%%%%%%%%%%%%%%%%%%%%%%%%%%%%%%%%%%%
%%%%%%%%%%%%%%%%%%%%%%%%%%%%%%%%%%%%%%%%%%%%%%%%%%%%%%%%%%%%%%%%%%%%%%%%%%%%%%
%%%%%%%%%%%%%%%%%%%%%%%%%%%%%%%%%%%%%%%%%%%%%%%%%%%%%%%%%%%%%%%%%%%%%%%%%%%%%%

As noted in Section \ref{abrageo}, in the geometrical macroscopic picture, there is a one-to-one correspondence 
between impact parameter, spectator mass and charge, and excitation energy ($A$, $Z$, $E^*$). In the present work, 
for the LSD-Lang combination, we consider a few prefragments, only. It is hoped that the restriction to 
some cases permits to better understand which region of the initial ($A$, $Z$, $E^*$) phase space contributes to 
a specific region of the populated final products. The selected nuclei are marked with  black crosses in 
Fig.~\ref{fig.02}:  $^{194}$Os, $^{181}$Lu, $^{168}$Dy, $^{152}$Nd, $^{142}$Ba, $^{128}$Sn, $^{100}$Y, and $^{64}$Mn, 
formed in collisions with impact parameter decreasing from around 12 to 4 fm. The initial excitation energy is below 150 MeV ('sphere-cylinder' scenario), 
increasing roughly linearly with increasing abraded mass, see Fig.~\ref{fig.02} (b). 

\begin{figure}[!hbt]
   \begin{center}
     \includegraphics[scale=0.70]{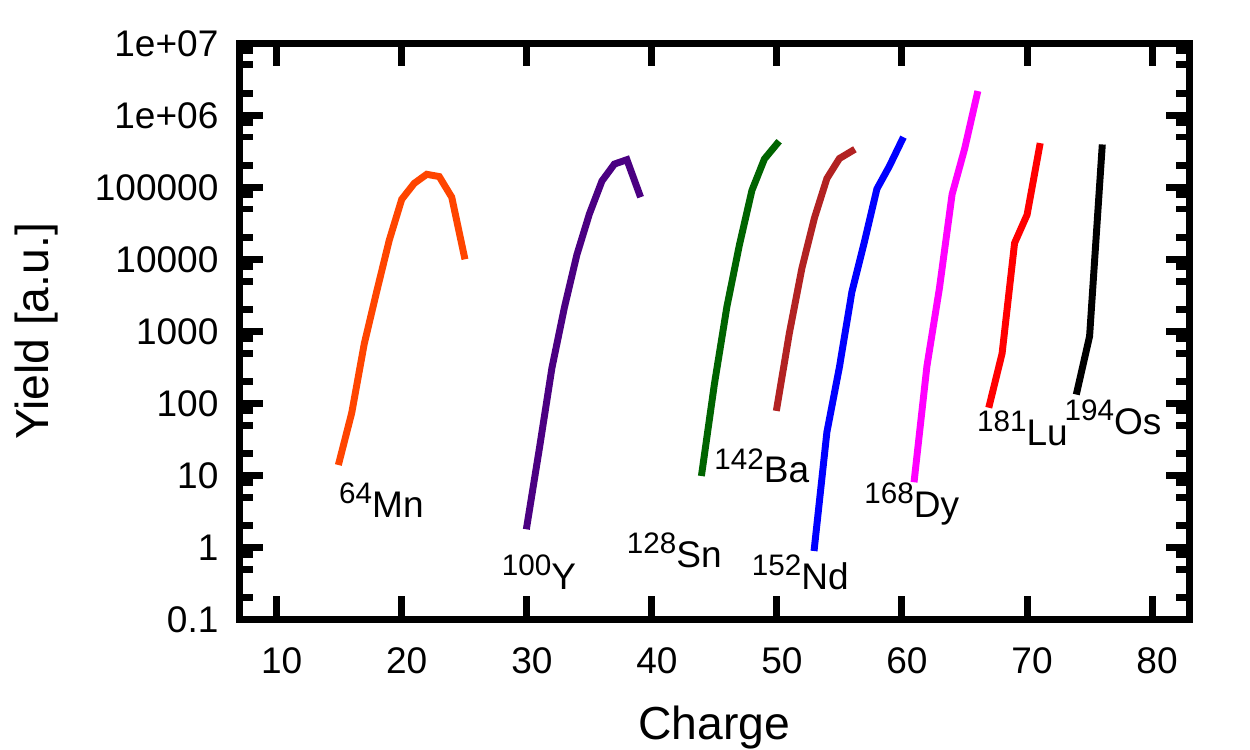}
     \caption{(Color on-line) Charge distribution of the evaporation
       residues produced along the decay of the set 
     of prefragments selected for the LSD-Lang calculation. The spectator is specified next to each curve. The integral of each distribution was scaled according to the probability that the decay ends in the ER channel.
             }
                                                                 \label{fig.06}
   \end{center}
\end{figure}
\begin{figure}[!hbt]
   \begin{center}
     \includegraphics[scale=0.60]{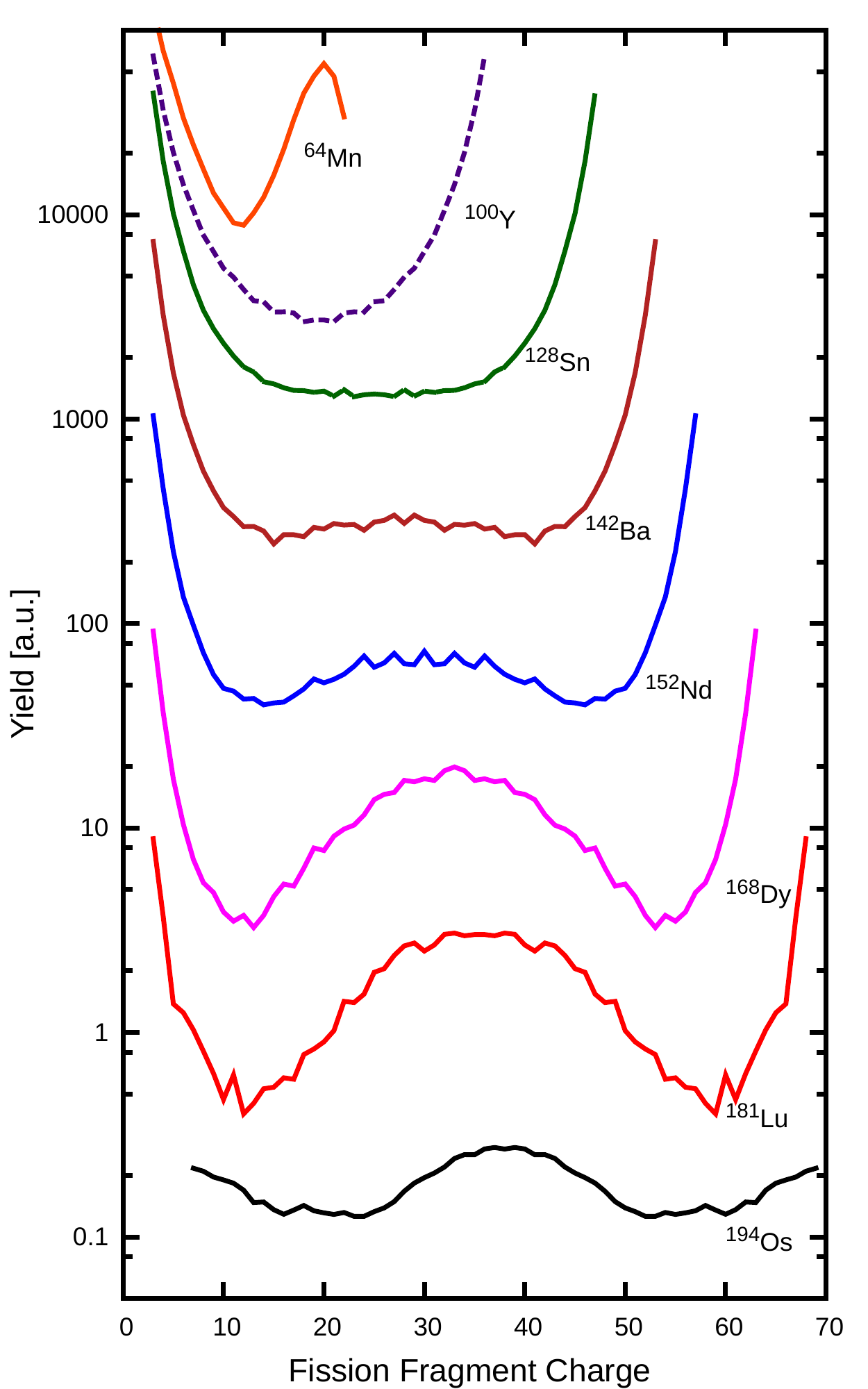}
     \caption{(Color on-line) Charge distribution of the fission fragments  produced along the decay of the set of prefragments 
selected for the LSD-Lang calculation.  The spectator is specified next to each curve. The integral of the curves 
do not reflect the fission probabilities (which are given in Table~\ref{tab01}); curves were displaced arbitrarily 
in the vertical direction for better legibility.
             }
                                                                 \label{fig.07}
   \end{center}
\end{figure}

\begin{figure*}[!bt]
   \begin{center}
     \includegraphics[scale=1.10]{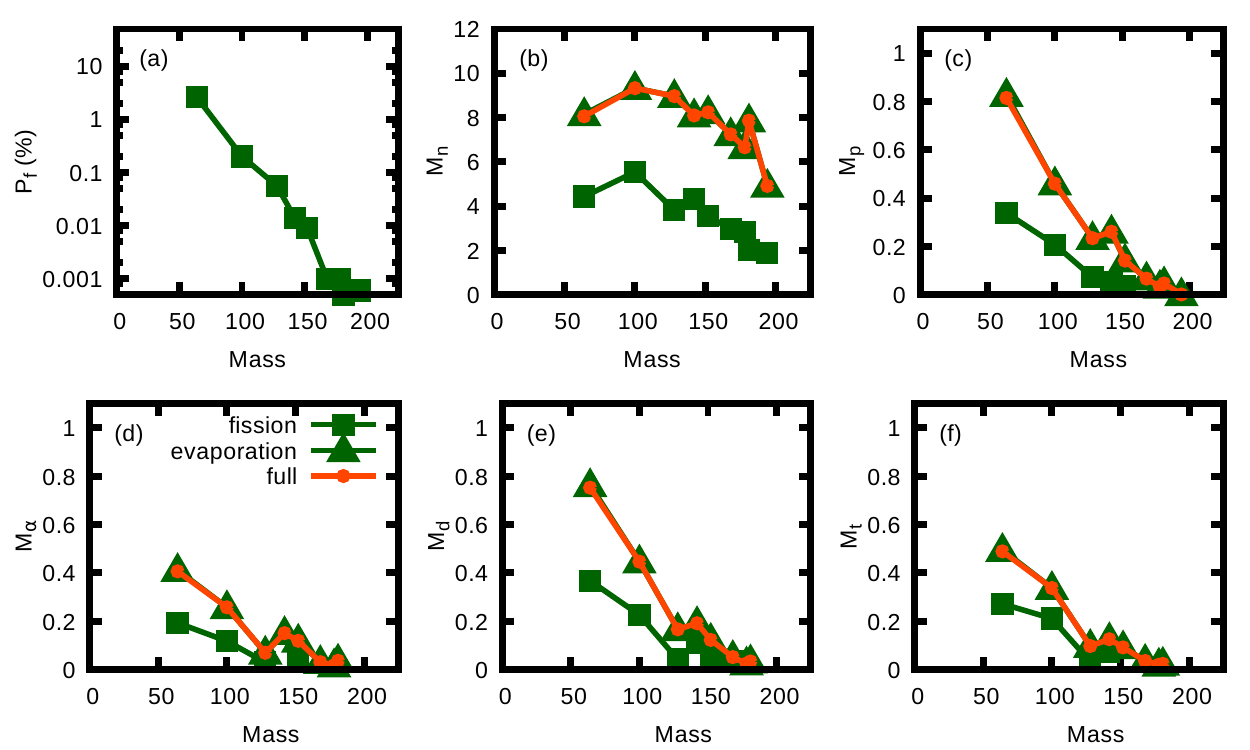}
     \caption{(Color on-line) Fission probability and mean
       multiplicities of neutrons, protons, $\alpha$ particles, 
     deuterons and tritons emitted along the deexcitation of the excited
     spectator on its way to either formation of an 
     evaporation residue (green triangles), or fission (green squares). 
     The average multiplicities, irrespective of
     the decay channel, are shown also (orange dots). }
                                                                 \label{fig.08}
   \end{center}
\end{figure*}

Figure~\ref{fig.06} presents the charge distribution of the ERs produced along the decay of the considered prefragments. 
The integral under the curves reflects the ER channel probability; it decreases with increasing abraded mass due to the 
increasing excitation energy and thus enhanced fission probability. However, the magnitude of the difference is very 
small, since ER decay largely dominates (see further below). Note that weighting by the impact parameter distribution 
is not included at this step. With increasing abraded mass, the maximum of the ER distribution is observed to shift away from the charge of the prefragment, leading to a progressive change in the shape of the distribution. This is, of course, due to the increased probability of charged-particle evaporation with increasing prefragment $E^*$ and, to lesser extend, decreasing mass.

The charge distribution populated in the fission channel (complementary to ER) is shown in Fig.~\ref{fig.07}, again for 
the sample of selected prefragments. Note that, here, the presented yields do not reflect the fission probability; curves 
were arbitrarily displaced vertically for clarity. While fission of the heaviest prefragments is symmetric (centered around 
about half the charge of the prefragment), the fission partition becomes progressively asymmetric (the two fragments are not 
of equal charge) with increasing abraded mass. The shape of the distributions roughly reflect the topography of the energy 
landscapes presented in Fig.~\ref{fig.05}:  for $^{100}$Y with a mean temperature at scission $<T_{sc}>$=2.91 MeV, $^{142}$Ba with 
$<T_{sc}>$=2.18 MeV, $^{168}$Dy with $<T_{sc}>$=1.73 MeV, and $^{194}$Os with $<T_{sc}>$=1.17 MeV. The lower the energy, 
the more favored the corresponding $q_3$ partition. This change in the shape of the fragment charge (equivalently, mass) 
distribution illustrates the signature of the Businaro-Gallone (BG) transition, located between Sn and Ba in the present 
temperature regime  \cite{mazurek:2016} for the LSD model.\\  

A summary of the properties of the prefragment decay is given in  Fig.~\ref{fig.08}: the fission probability and number 
of particles ($n$, $p$, $\alpha$, $d$, $t$) emitted along the decay in the ER and fission channels are displayed as a 
function of spectator mass. The increase of the fission probability with increasing abraded mass is due to the corresponding 
increase in prefragment excitation energy. Though, it is to be noted that the fission probability remains small in all 
cases (below a few $\%$) due to the low fissility of nuclei situated below Pb. Independent of the decay channel 
(ER or fission) the multiplicities of the emitted particles also increases from $^{194}$Os to $^{64}$Mn, reflecting again 
the increasing excitation of the product left after abrasion. The variation of the neutron multiplicity (panel b) tends though 
to develop a plateau with decreasing spectator mass. This is related to the increasing competition of charged-particle 
emission (panels c to f) for higher excitation and lighter systems, and
which takes away a part of the energy available for
 neutron evaporation. Finally, it is observed that more particles (of any kind) are emitted in the ER channel than in the 
 fission channel. The reason behind this difference is the energy required by the system to overcome the (large) fission 
 barrier, which energy is then not available any more for evaporation.\\
We show also in Fig.~\ref{fig.08} the average multiplicities independent of the decay channel. They are, of course, very 
close to those obtained when selecting the ER channel, since the latter widely dominates. It 
is customary in theoretical work in the low energy domain (below about $10$ MeV/nucleon) to analyze the 
ER and fission channels separately, since they are usually tagged in experiment. However, in studies at 
ultra-relativistic energies, experimental information about the decay of the spectator is very scarce as 
discussed above. In addition, according to the high velocity-boost of the projectile, all products (heavy and light) are 
strongly forward focused and moving fast. Discriminating particles from the ER and fission channels is therefore very 
difficult, and was not attempted yet to our knowledge. Hence, it is the average multiplicity, irrespective of the 
fate of the spectator, that is most useful for comparison with experiment 
on particle multiplicities wherever available.        

%
%--------------------------------------------------
%\subsection{Excitation energy studies}
\subsection{Gaimard-Schmidt abrasion coupled to dynamical decay and comparison with ABRABLA}
%--------------------------------------------------

As noted earlier, the output of the abrasion model based of Refs.~\cite{gaimard:1991, schmidt:1993}, in terms of prefragment mass, charge, 
excitation energy and angular momentum, is used as input for the dynamical code, and is referred to as ABLA-Lang. Combining the predictions 
by the LSD-Lang and ABRA-Lang softwares will permit to shed light on the influence of the first stage of the reaction. The predictions 
are further compared to the results of the ABRABLA code, yielding information, in this case, about the second stage.

\begin{figure}[!bt]
   \begin{center}
     \includegraphics[scale=0.990]{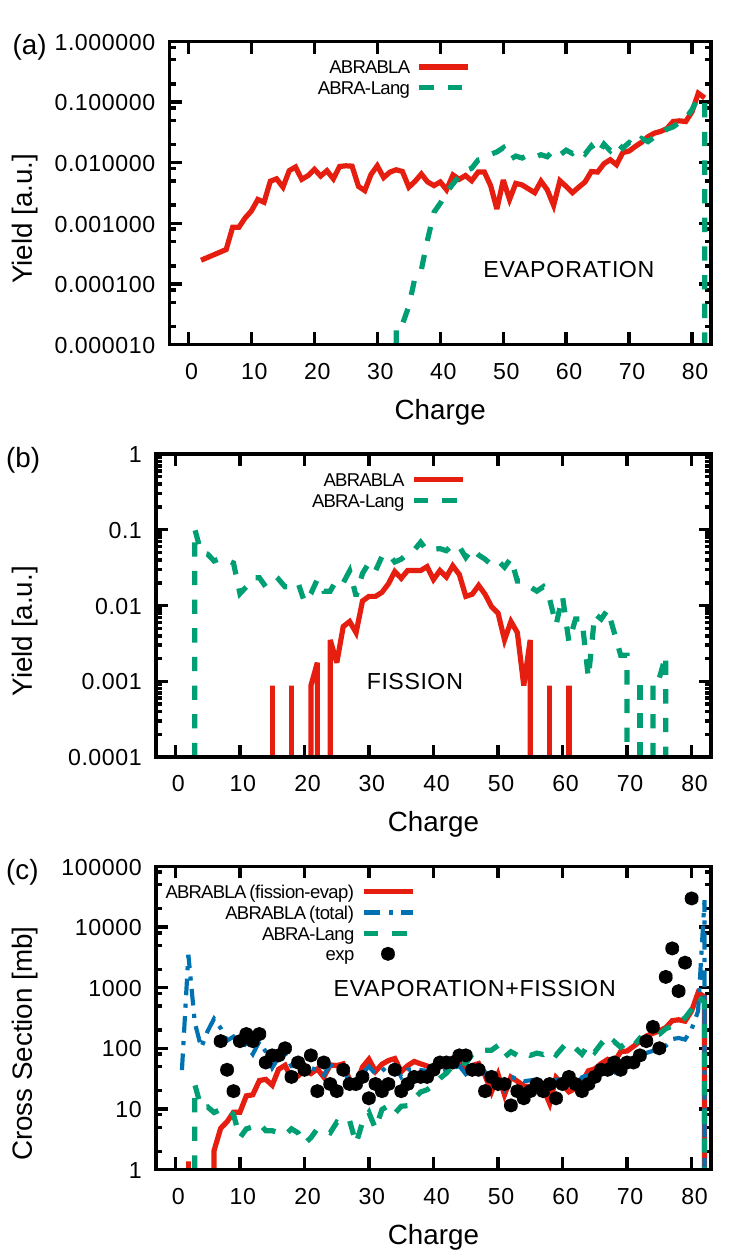}
    \caption{(Color on-line) Charge distribution of the evaporation residues (a) and fission fragments (b) obtained 
    with the ABRA-Lang (dashed line) and ABRABLA (full line) codes. Weighting by the impact parameter distribution predicted 
    by ABRA is performed in both cases. The integral distributions including ER and fission channels are 
    shown in (c); normalization to absolute cross sections is performed using the nuclear-induced reaction cross section extracted 
    in experiment \cite{cecchini:2002}. The total distribution predicted by the ABRABLA code, with no event selection 
    (that is, including electromagnetic-induced interactions, and break-up processes) is displayed as well with the dash-dotted 
    curve; absolute normalization is directly taken from the total reaction cross section predicted by ABRABLA. Experimental 
    data are from Refs.~\cite{dekhissi:2000,cecchini:2002}.
%    \color{red}{I still have much difficulty to understand 
%    the panel (a). Also the normalization of (b) is not understood by me} \color{black}{}.
             }
                                                                 \label{fig.09}
   \end{center}
\end{figure}
The outcome of ABRA-Lang and ABRABLA is gathered in Fig.~\ref{fig.09} for the charge distribution for which 
 experimental data exist for \cite{dekhissi:2000,cecchini:2002} for the
 $^{208}$Pb (158 GeV/nucleon) +$^{208}$Pb collision. 
 Figures~\ref{fig.09} (a) and (b) restrict, respectively, to the ER and fission channel, while (c) includes both channels. Note 
 that, since ABRA-Lang uses the input file produced by ABRA, it automatically considers the same impact parameter distribution as 
 ABRABLA. The ER charge is observed to extend to much smaller values in ABRABLA as compared to ABRA-Lang. 
 On the contrary, the fission-fragment charge distribution scans a wider domain for ABRA-Lang than for ABRABLA. The
 fission-fragment charge distribution is Gaussian-like and well localized for ABRABLA, suggesting that fission is mainly 
 populated by the decay of heavy prefragments with $Z$ from about 60 to 80. The heaviest, lowly excited, and lightest, 
 very excited, prefragments experience, respectively, very short and long evaporation cascades. 
 On the other hand, in ABRA-Lang, the 
 light prefragments at the high excitation predicted by ABRA undergo fission. This explains the non-symmetric shape of the ABRA-Lang 
 distribution in Fig.~\ref{fig.09} (b) which results from the convolution of the individual distributions of Fig.~\ref{fig.07}. 

\begin{table}[h]
\caption{ER and fission observables as predicted by the LSD-Lang ('sphere-cylinder' scenario), 
ABRA-Lang and ABRABLA codes for 
$^{208}$Pb (158 GeV/nucleon) +$^{208}$Pb. All calculations use the impact parameter distribution predicted by ABRA, and
averaging is performed over all products in the ER (respectively, fission) channel. The listed quantities 
correspond to the fission probability $P_f$, number of light particles emitted in each channel: 
multiplicities of pre-scission neutrons, protons and $\alpha$'s (n$_{pre}$, p$_{pre}$, $\alpha_{pre}$), and of the
particles leading to a cold ER (n$_{ER}$, p$_{ER}$, $\alpha_{ER}$). The fission time $t_f$ is 
given also. 
%\color{red}{(You do not give the results for alpha's for ABRABLA. I suspect that 
%this is because, as I told you, it is not possible to extract easily multiplicities from the standard ABRABLA output. You 
%did in some way, which I ignore. That may also explain why we are not able to understand the results for ABRABLA for the 
%n and p, as they may be ring because alpha are not accounted for ??)} \color{black}{}
}
\begin{tabular}{ccc}\\
\hline\\
%&\multicolumn{2}{c}{FRLDM}&\multicolumn{2}{c}{LSD}&Exp.\\\\
%\hline   
                       &LSD-Lang &ABRA-Lang \\
\hline\\
$P_f$  			& 0.00014 &  0.01879 \\
n$_{pre}$       	& 3.0035  &  15.0651 \\
n$_{ER}$       		& 6.9075  &  16.264  \\
p$_{pre}$       	& 0.0208  &  2.586  \\
p$_{ER}$       		& 0.0953  &  1.9427 \\
$\alpha_{pre}$       	& 0.0099  &  0.590  \\
$\alpha_{ER}$      	& 0.0530  &  0.572   \\
%$<M>$,  a.m.u.         & 82.051  &  75.881 & 94.5\\
%$<Z>$,  a.m.u.         & 32.802  & 31.858 & 40 \\
%$<TKE>$, MeV         	& 65.917  & 81.901 &  \\
%T, MeV	               & 1.7628  &  3.627 & \\
t$_f$, $10^{-21}$s	 & 11.272 & 14.636  \\\\
%$<E^*_{sc}>$, MeV	& 60    &   63  &   74  &   71  &  \\
\hline
\label{tab01}
\end{tabular}
\end{table}

The comparison between various observables commonly investigated in the field is given in Table~\ref{tab01}. The average fission probability presented in Table~\ref{tab01} shows indeed a huge increase between LSD-Lang and ABRA-Lang, due to the 
 larger prefragment spectator excitation. 
 %Though, the average fission probability is seen to be higher for ABRABLA as compared to ABRA-Lang. 
 %This is understood by the fact that the value for ABRABLA is the average involving mainly only heavy prefragments, whereas 
 %the probability for ABRA-Lang includes also lighter systems \color{red}{(Still not clear with this at all)} \color{black}{}. 
 These 
 observations shows that the Langevin code predicts more fission in light systems, than the statistical 
 ABLA model. One reason for explaining this difference may be the parameterization of the empirical potential used in ABLA, adjusted 
 to fission of heavy nuclei, and which may not be best-suited any more around and below the BG transition. Another reason of the 
 difference may be related to dissipation effects. Elucidating the intricate interplay of these (and others) possible 
 reasons is beyond the scope of this work.\\
The substantial difference between 
 the light-particle multiplicities predicted by LSD-Lang and ABRA-Lang directly reflects the difference in initial excitation energy
 as depending on the abrasion model. A larger fission time for ABRA-Lang is connected to the time required to emit more particles 
 before fission. 
 %The difference between the values of ABRA-Lang and ABRABLA is less pronounced, and may result from the subtle interplay 
 %mentioned above. 
 It is not easy to trace back at this stage, as it additionally includes an averaging over prefragments with different 
 yields in each channel, depending on the model, as discussed above around Fig.~\ref{fig.09} (b).\\
 Finally, in Fig.~\ref{fig.09} (c) we display the sum of the ER and fission channels as predicted by ABRA-Lang and ABRABLA. Normalization 
 to absolute cross sections is performed with the nuclear-induced reaction cross section extracted in experiment \cite{cecchini:2002}. The
 rather remarkable description of the shape of the experimental distribution with ABRABLA (red full line) in the region where the 
 ER and fission dominate is noteworthy. The description by ABRA-Lang (green dashed line) is very encouraging, having in mind its  
 dynamical framework, which is confronted to such kind of data in a 'brute-force' manner here for the first time.
 
 We overlay also in Fig.~\ref{fig.09} (c) the prediction by the ABRABLA
 code (blue dash-dotted line), including all types of interactions (nuclear- 
 and electromagnetic-induced), all kinds of decay channels (ER, fission, with or without break-up, IMF emission), and normalized with 
 the ABRABLA-predicted total reaction cross section. Inclusion of electromagnetic-induced reactions improves the description close to 
 the projectile mass, as expected, while break-up and IMF contribute to enhance the yields of lighter products. The overall description 
 of ABRABLA is very good. So far, the achievement of the code was studied in detail and demonstrated powerful at relativistic beam 
 energy. To our knowledge, this is the first time that it is tested at ultra-relativistic energy, showing good extrapolation 
 property.

\subsection{Reaction time scales and possible connection with pion
  electromagnetic effects}

%\color{red}{(I again need to discuss the goal here - Please check VERY carefully)} \color{black}{}
%\\

\begin{figure}[!bt]
   \begin{center}
     \includegraphics[scale=0.60]{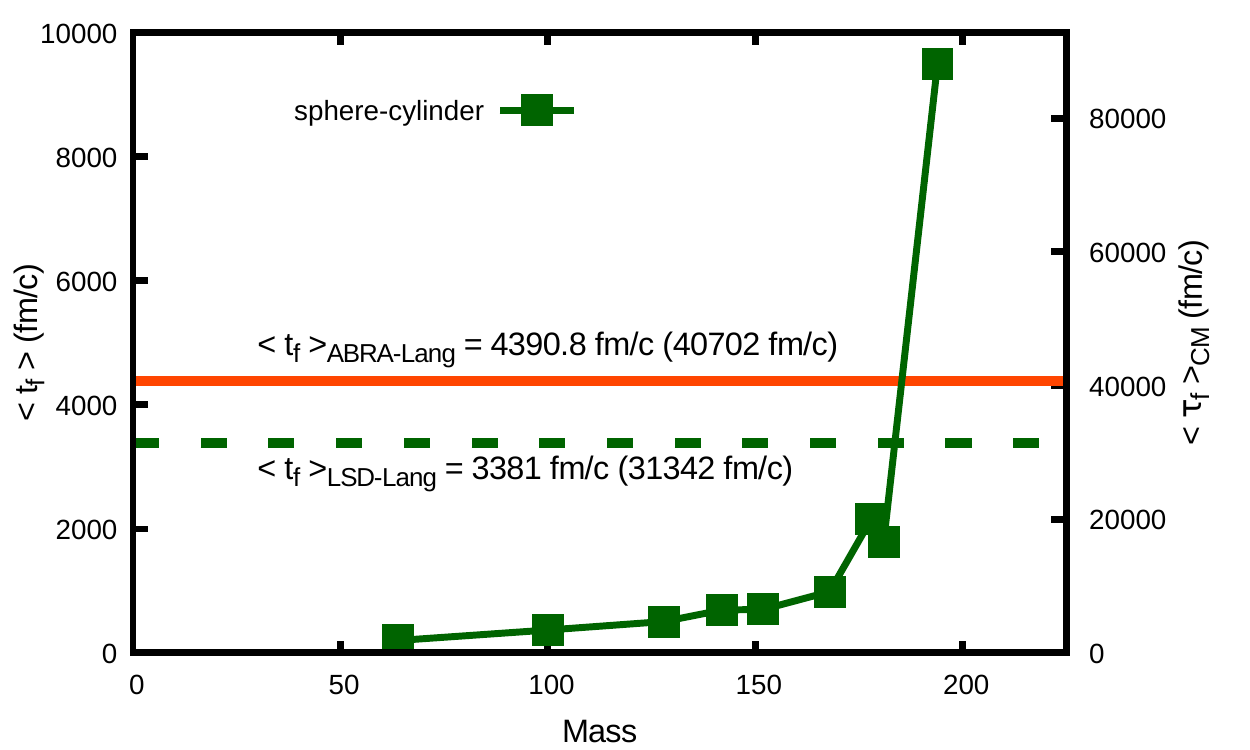}
    \caption{Mean fission time as a function of spectator mass in the LSD-Lang code. The right-side $y$-axis gives 
    the time as output from the Langevin calculation, in the fission frame, and which is transformed to the overall
    center-of-mass time on the left-side $y$-axis. Horizontal lines indicate the mean fission time averaged over all 
    spectators for the LSD-Lang (green dashed) and ABRA-Lang (orange full) codes. 
%    \color{red}{(Are the average times weighted by impact parameter  distribution ?)} \color{black}{} 
             }
                                                                 \label{fig.10}
   \end{center}
\end{figure}

In Ref.~\cite{rybicki:2007} it is proposed that the distortion observed in the positively- and negatively-charged 
pion spectra is caused by the electromagnetic field generated by the heavy spectators 
moving at relativistic velocity, provided that these live long enough. In this context, one of the main assets of dynamical 
calculations used in this work is the possibility to predict the decay time of the excited spectator, including the 
time scale (and sequence) of the light particles emitted along its decay, {\ it i.e.} the time taken to either reach a cold ER or 
to fission into two fragments.\\
As noted earlier, the main decay channel of the excited spectators
formed in lead on lead collisions is found to be the formation of a heavy residue. 
From Table~\ref{tab01} we see that the corresponding events are characterized by emission of neutrons mainly\footnote{The charge of the ER
is in 90$\%$ of the cases above 70, independent of the models used (note the logarithmic scale in Fig.~\ref{fig.09} (a)}). That is, the collision 
leads to two highly-charged residues, in the vicinity of the pions
produced by fireball or from fire-streak at the SPS energies. These
residues live long enough to interact electromagnetically with pions. 
%\color{red}{(forever actually....) do we need to 
%indicate the formation time of the cold ER ??} \color{black}{} . 
For the remaining events, {\it i.e.} when fission occurs, much 
smaller product charges are reached. That can reduce electromagnetic
effects on pions, except when the system lives long-enough before splitting into 
two parts. The Langevin approach is particularly suited to investigate
the fission time scale. 
In Fig.~\ref{fig.10} the mean fission time is displayed as a function of
prefragment mass in the rest frame of the spectator system\footnote{The corresponding time in the overall
  center-of-mass system is larger due to Lorentz dilatation by about a
  factor of 10, see left-side $y$-axis.}, 
as obtained with the LSD-Lang code 
  ('sphere-cylinder' scenario). Longer times for the heavier prefragments are due to the smaller excitation energy in 
  peripheral collisions. Also given is the 
  mean time averaged over all spectators (horizontal lines) as obtained with the LSD-Lang and ABRA-Lang ($\mathrm{<t_f>_{LSD-Lang}}$ 
  and $\mathrm{<t_f>_{ABRA-Lang}}$, respectively). 
The predicted fission time in the rest-frame of the spectator is in the range 50-3000~fm/c, and even longer for very 
peripheral collisions. After Lorentz transformation, this yields an overall center-of-mass time $\tau_{\rm CM}$ of the colliding nuclei which 
is enough to guarantee long-lasting electromagnetic interactions between the spectators and the pions ejected from
the quark-gluon plasma. In the numerical simulations of the electromagnetic effects 
the trajectories are followed till $\tau_{\rm CM}\sim$~1000~fm/c.
\\

%\color{red}{(LOST LOST LOST)}
The time measured in the spectator reference frame can be transformed to the overall
center-of-mass system (right $y$-scale in  Fig.~\ref{fig.10}) as:
\begin{equation}
\tau_{CM} = \frac{t_{spec}}{\sqrt{1 - \beta_{spec}^2}} \; ,
\label{time_CM}
\end{equation}
where $\beta_{spec}$ is the relativistic velocity of the spectator
in the overall center-of-mass system.
We assume: $\beta{spec} = \beta_{A_i/CM}$, where the latter
is velocity of the projectile/target in the overall CM system.
Then:
\begin{equation}
\beta_{spec} = \frac{\sqrt{s/4 - m_N^2}}{\sqrt{s}/2}
\; .
\label{beta_spectator}
\end{equation}

For the maximal NA49 energy $\sqrt{s_{NN}}$ = 17.3 GeV we get
$\beta_{spec}$ = 0.9942 and the dilatation factor is of about 10. The spectators could live even
more than 40000~fm/c.

Thus, we conclude that the spectator systems, whatever the final fate is (ER or fission), live long enough to cause
the electromagnetic effect observed in \cite{rybicki:2004,rybicki:2011}.

In the experiments such as NA49 or NA61 the colliding heavy ions have energy  
2 GeV~$\leq\sqrt{s_{NN}}\leq$~17.3~GeV. The experiments at SPS
concentrate on measuring rapidity 
and transverse momentum distributions of pions, kaons,
nucleons, etc. They can also measure residues of the collisions in very
forward direction 
(see e.g. \cite{dekhissi:2000,cecchini:2002}). So far such measurements
are done independently. We do not know whether they can be done in
coincidence which would provide new information on how the 
participant and spectator systems are correlated. Also, it could enlight the predictions of this work.

% \color{black}{}
 
%\clearpage
%--------------------------------------------------------------------------
\subsection{Decay at specific impact parameter}
%--------------------------------------------------------------------------
%------------------------------------------------
%\section{Plans for the current paper, to do list}
%------------------------------------------------

%1) Concentrate on $b =$ 10.5 fm relevant for the
%   electromagnetic effect observed by NA49.

%   Calculate excitation energy from \\
%   (a) geometry+LDM \\
%   (b) ABRABLA code \\
%   (c) Formulas given in the theory section on AA (abrasion-ablation) model \\

%   Calculate decay with the Langevin approach \\
%   Get times of the decay   

%2) Calculate $Z$ distribution for Pb+Pb and compare with
%   the NA49 data. (!!!!!!!!!!!111)

%------------------------------------------------------------

The investigation of Ref.~\cite{rybicki:2007} about the influence of the spectator electromagnetic field on 
pion trajectories was performed at fixed impact parameter $b \approx$~10.5 fm. In this paragraph we therefore sort 
the calculations according to impact parameter. In practice, we do restrict to those predictions corresponding
to $b \in$ (10-11)~fm.

The outcome of the first abrasion stage, in terms of prefragment properties, is 
very different depending on the abrasion model used. In the geometrical macroscopic picture ('sphere-cylinder scenario'),
the prefragment predicted in this impact parameter slice is peaked around $^{181}$Lu with an excitation energy slightly 
less than 100~MeV. For the Glauber model, very few nucleons are removed: the remnant is sharply centered around 
$^{206}$Pb with $E^* \in$ (20-50)~MeV. Finally, according to ABRA, the prefragment mass is characterized by a 
wider distribution around $A \approx$~170, similar to the geometrical macroscopic approach, but with
a much larger ($E^*$ above 500~MeV).

%...............sigma(N)...for the above range of impact parameter........
%\begin{figure}[!bt]
%   \begin{center}
%     \includegraphics[scale=0.60]{glauber_abra_A_distr_b105.pdf}
%     \includegraphics[scale=0.60]{glauber_abra_Eexc1020_distra.pdf}
%    \caption{The distribution of the spectator mass (a) and excitation energies 
%    obtained with Eq.~\ref{average_excitation energy} (b) for impact parameters 
%    $b \in$ (10,11)~fm. 
%             }
%                                                                 \label{fig.11}
%   \end{center}
%\end{figure}

\begin{figure}[!bt]
   \begin{center}
     \includegraphics[scale=0.60]{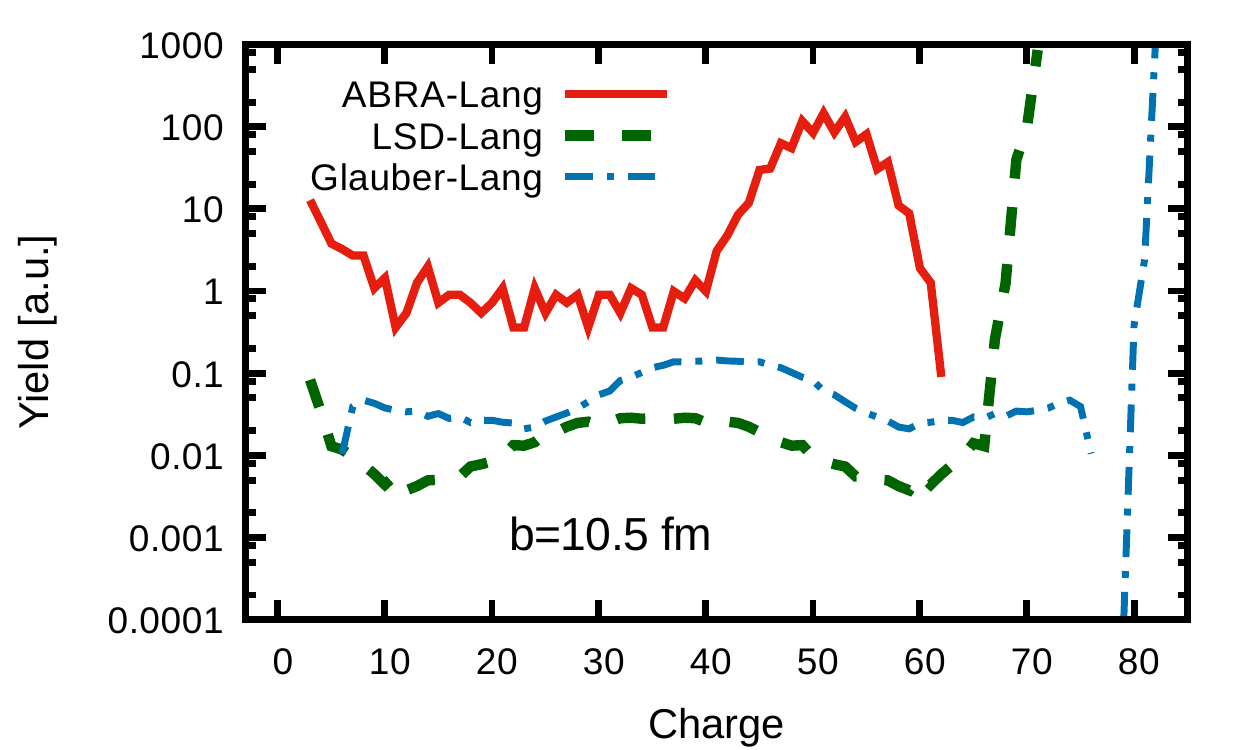}
    \caption{(Color on-line) Charge distribution of the products (ER and fission fragments) predicted with three abrasion 
    models combined with the same Langevin decay code, for the restricted range of impact parameter $b \in$ (10-11)~fm. 
%    \color{red}{(From the text, ABRABLA should read ABRA-Lang ?? But in that case, I do not understand how ABRA-lang can give 
%    this red curve. I suspect that it is indeed ABRABLA then ?? And the bump is the ER after a lot of evap ?)} \color{black}{}
    }
                                                                 \label{fig.12}
   \end{center}
\end{figure}

%\color{red}{(I propose to remove fig 13 (in previous version) for two reasons: 1) the single point for LSD is difficult to see (
%by the way , $b$ is not strictly fixed, so I would have expected a couple of values, a narrow distribution, as suggested by 
%e.g. figure 1, and
%2) the difference in A like in $E^*$ is SO big that the reader may wonder what all this study means: That is: differences are so huge 
%all along that the reader may say that we should be clever enough to estimate that one or the other picture is less reliable. Otherwise,
%it means that that we do not understand physics. Saying in the text what is in the figure, is well enough to me... In case 
%you decide to keep, I presume that the legend ABRABLA should be ABRA ?)} \color{black}{}

%\color{red}{(I propose also to remove the discussion around Glauber predictions. First, we "judge" the model too strongly, and 
%finally say that we do not know why it is not good (easy to say...). In addition, why should any other model be better ? We do not
%know neither. Finally, the discussion around Glauber does not bring anything for the text which follow in the last part of the
%chapter).} \color{black}{}
 
%\color{red}{(I also removed the discussion around ABRABLA and break up and so on. The reader may wonder again about this huge 
%difference between abrasion models, and whether what we do makes sense at all).} \color{black}{}
%\\
%\\
The outcome of the three abrasion models were combined with the Langevin code in order to compute the decay of the hot prefragments, as 
explained in the previous section, and then sorted according to $b \in$ (10-11)~fm. The 
resulting distributions for the final product charge are shown in Fig.~\ref{fig.12}. Whereas ER dominate the distribution for LSD-Lang and 
Glauber-Lang, fission is the main decay channel for ABRA-Lang, due to the much higher excitation energy involved. The fission time 
amounts to $\tau_{spec} \sim$1792~fm/c, $\sim$471~fm/c, and $\sim$711273~fm/c for LSD-Lang, ABRA-Lang and 
Glauber-Lang, respectively. All these times are sufficiently large for the electromagnetic effects predicted in 
Ref.~\cite{rybicki:2007} to be active and affect the trajectory of the
pions emitted from the fireball. That is, for those events which do not
produce a final heavily-charged ER, and for which fission instead takes place, the decay by fission is slow enough for having at hand a still very
heavy charge for quite some time and which influences the pion
trajectories and momentum distributions.    
%--------------------------------------------------------------------

%%%%%%%%%%%%%%%%%%%%%%%%%%%%%%%%%%%%%%%%%%%%%%%%%%%%%%%%%%%%%%%%%%%%%%%%%%%%%%%%%%%%%%%%%%%%%%%%%%

%--------------------
\section{Conclusions}
%--------------------

The decay of the heavy excited spectator remnants produced in peripheral heavy-ion 
collisions at ultra-relativistic energies, and which has received a poor attention so far, 
is investigated with an innovative theoretical framework. Within the picture of the well-established two-stage reaction scenario, 
three abrasion models - borrowed from the relativistic beam energy domain, were combined with a dynamical 
model based on the stochastic Langevin 
approach - popular in the physics at Coulomb barrier energies for modeling the competition between 
evaporation and fission. Statistical-model 
calculations with the ABRABLA code, which software has shown very powerful for collisions of
relativistic heavy ions, are considered as well.\\
Comparison between various model combinations for the first and second stage of the reaction allows to
study the influence of the predicted prefragment spectator mass and excitation energy after abrasion, as well as
the influence of the modeling of the excited prefragment decay, on the final product charge distribution. The ABRABLA
code is observed to describe the corresponding available experimental data rather well, demonstrating its good extrapolation properties in the ultra-relativistic domain where it was never tested. Also, the dynamical calculation computed within the Langevin approach, and combined with a reasonable abrasion model, shows a rather promising tool for modeling the decay of the heaviest 
spectators produced in peripheral collisions.\\
The main asset of the here-proposed Langevin approach lies in the possibility to predict the time evolution of the 
spectator, in contrast to purely statistical codes. Recent theoretical studies suggest that understanding 
this evolution may be crucial to consistently explain the pion spectra observed in the energy domain typical of the 
future CERN SPS or RHIC facilities. The time scale for the decay of the spectator predicted with our new theoretical 
framework within this field is consistent with the presence of a heavy system that lives long enough to impact the 
trajectory of the pions from the fireball region by its strong electromagnetic field.\\
        
The present work shows the widespread potential for the implementation of the stochastic Langevin approach. At Coulomb 
barrier energies, the method was successfully used to get insight into nuclear dynamics, and more specifically friction of 
nuclear matter \cite{nadtochy:2012,adeev:2005,mazurek:2017}. The approach also demonstrated to 
be a pre-requisite tool for un-ambiguously understanding the dynamics and subtle time evolution of nuclei across the Businaro-Gallone 
transition \cite{mazurek:2016}. This exploratory work suggests that it can be an interesting approach for the 
ultra-relativistic energy community for describing the fate of peripheral collisions. Beside the aspect of pion trajectory 
mentioned in this work, combining such studies with work on abrasion-induced 
reactions at relativistic energy may also be a relevant playground to investigate the disappearance of friction effects in the 
entrance channel of the reaction with increasing beam energy. The latter is expected to lead to lower excitation energy of the 
spectator, which would surely affect its time evolution and decay.   

%The signal of such a transition would be a sudden drop of neutron
%multiplicity with increasing collision energy.
%This could be measured by the NA61 experiment at CERN or at RHIC with
%zero-degree calorimeter(s).

%\color{red}{(Finally, it seems to be that we are not so sensitive to the spectator E* vs. A correlation so much:
%I mean: the different abrasion models give very different (A,E*), but finally all give very similar Z distributions and 
%reasonably well when compared to expt ? Is this poor sensitivity it because we have a mixture (big soup!) of so many 
%prefragments, that finally we are not sensitive to anything  ??? Or, is it because we 
%restrict to exptal Z distributions ? Isotopic and mass would be more discriminant ?? I am not convinced. 
%Does all this maybe just mean that most important is the distribution of impact parameter ??)} \color{black}{}

\vskip 1cm

{\bf Acknowledgements}
\\
This work was partially supported by the Polish National Science Centre under Contract No. 2013/08/M/ST2/00257 
(LEA COPIGAL) (Project No.~18) and IN2P3-COPIN (Projects No.~12-145, 09-146), and by the Russian Foundation for 
Basic Research (Project No.~13-02-00168). We are very thankful to Dr. Aleksandra Kelic-Heil for calculations 
with the ABRABLA code and related analysis. We are also indebted to Andrzej Rybicki for discussion about experiments 
at the CERN SPS.
%......................
%%%%%%%%%%%%%%%%%%%%%%%%%%%%%%%%%%%%%%%%%%%%%%%%%%%%%%%%%%%%%%%%%%%%%%%%%%%%%%%%%%%%%%%%%%%%%%%%%%
%%%%%%%%%%%%%%%%%%%%%%%%%%%%%%%%%%%%%%%%%%%%%%%%%%%%%%%%%%%%%%%%%%%%%%%%%%%%%%%%%%%%%%%%%%%%%%%%%%
%%%%%%%%%%%%%%%%%%%%%%%%%%%%%%%%%%%%%%%%%%%%%%%%%%%%%%%%%%%%%%%%%%%%%%%%%%%%%%%%%%%%%%%%%%%%%%%%%%
%%%%%%%%%%%%%%%%%%%%%%%%%%%%%%%%%%%%%%%%%%%%%%%%%%%%%%%%%%%%%%%%%%%%%%%%%%%%%%%%%%%%%%%%%%%%%%%%%%
%%%%%%%%%%%%%%%%%%%%%%%%%%%%%%%%%%%%%%%%%%%%%%%%%%%%%%%%%%%%%%%%%%%%%%%%%%%%%%%%%%%%%%%%%%%%%%%%%%
%%%%%%%%%%%%%%%%%%%%%%%%%%%%%%%%%%%%%%%%%%%%%%%%%%%%%%%%%%%%%%%%%%%%%%%%%%%%%%
%%%%%%%%%%%%%%%%%%%%%%%%%%%%%%%%%%%%%%%%%%%%%%%%%%%%%%%%%%%%%%%%%%%%%%%%%%%%%%%%%%%%%%%%%%%%%%%%%%
\appendix
%\section{Appendix}
%%%%%%%%%%%%%%%%%%%%%%%%%%%%%%%%%%%%%%%%%%%%%%%%%%%%%%%%%%%%%%%%%%%%%%%%%%%%%%%%%%%%%%%%%%%%%%%%%%
%%%%%%%%%%%%%%%%%%%%%%%%%%%%%%%%%%%%%%%%%%%%%%%%%%%%%%%%%%%%%%%%%%%%%%%%%%%%%%%%%%%%%%%%%%%%%%%%%%
%\clearpage
\section{Surface and volume calculations}\label{a1}

\begin{itemize}
\item The volume of the sphere with radius $R_{sphere}=1.2A^{1/3}$ fm:
\begin{equation}
      V_{\rm sphere}(R_{sphere})
      =
      \frac{4}{3}\pi R_{sphere}^3,
                                                                \label{eqn.05} 
\end{equation}
with the surface of the sphere:
\begin{equation}
      S_{\rm sphere}(R_{sphere})
      =
      4\pi R_{sphere}^2.
                                                                \label{eqn.05} 
\end{equation}
\item The volume of the spherical cap reads:
\begin{equation}
      V_{\rm cap}(b)
      =
      \pi h^2(b) R_{sphere} - \frac{\pi}{3}h^3,
                                                                \label{eqn.05} 
\end{equation}
where $h(b)$ is the height of the cap; $a(b)$ - radius of the cap. $b$ - impact parameter.
The analytical formula for the surface of the spherical reads:
\begin{equation}
      S_{\rm cap}(b)
      =
      2 \pi h(b) R_{sphere}.
                                                                \label{eqn.05} 
\end{equation}
\item For the 'sphere-plane' (s-p) scenario the surface and the volume are:
\begin{eqnarray}
      S^{\rm s-p}(b) = S_{\rm sphere} -  S_{\rm cap}(b) +a^2(b)\pi,\\
      V^{\rm s-p}(b) = V_{\rm sphere} -  V_{\rm cap}(b).
\end{eqnarray}

\item For the 'sphere-sphere' (s-s) scenario the surface and the volume are:
\begin{eqnarray}
      S^{\rm s-s}(b) = S_{\rm sphere}, \\
      V^{\rm s-s}(b) = V_{\rm sphere} - 2 V_{\rm cap}(b).
\end{eqnarray}

The impact parameter is:
\begin{eqnarray}
      b^{\rm s-p}(b) = 2R_{\rm sphere} -h\\
      b^{\rm s-s}(b) = 2(R_{\rm sphere} -h)
\end{eqnarray}
\item The 'sphere-cylinder' (s-c) scenario is more complicated as there are no corresponding analytical formulas.

Our method to obtain the surface and volume of the spectator are presented below.

\begin{figure*}[!bt]
%   \begin{center}
\hspace{-1cm}
     \includegraphics[scale=0.90]{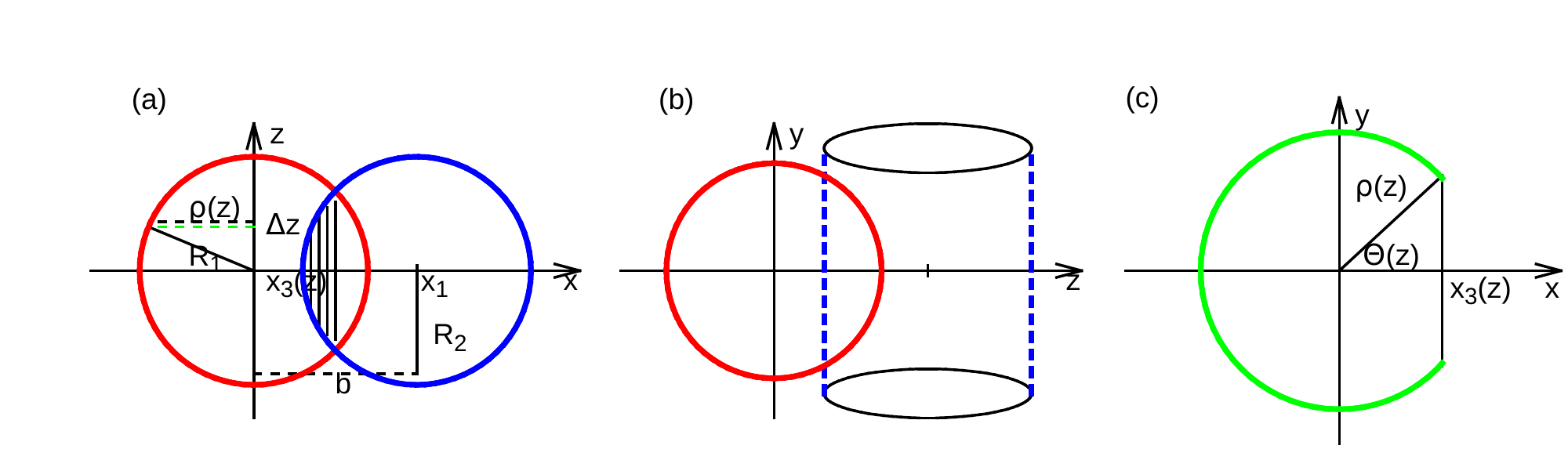}
     \caption{The cuts of the sphere in various planes. The red circle present the spectator-target, 
     blue-projectile, green- one layer in z-direction. 
             }
                                                                 \label{fig.A.02}
%   \end{center}
\end{figure*}

Let us consider characteristic points shown in Fig.~\ref{fig.A.02}:
\begin{eqnarray}
   && x_1=R_1+R_2 - b \quad {\rm (center\,\, of\,\, cylinder)},\\
   && x_2=x_1-R_2=R_1 - b  \\&&{\rm position\,\, of\,\,the \,\, inner\,\, surface\,\, for\,\, z=0},\nonumber\\
   && x_3=\frac{R_1^2-R_2^2+x_1^2}{2 x_1}, \quad \\&&{\rm position\,\, of\,\, extremal\,\, meeting\,\, point\,\, 'sphere-cylinder'}\nonumber\\
   && z_3=\sqrt{R_1^2-x_3^2} , \\&&{\rm position\,\, of\,\, crossing\,\, point\,\,of\,\, sphere\,\,and\,\,cylinder.}\nonumber
\end{eqnarray}
The deformed spectator is cut in the z direction with slices of $\Delta z$ long. In the plane (x,y) 
the shape of the slice looks like the circle without circular sector. The length of the chord, 
can be estimated with angle $\theta$
\begin{eqnarray}
 &&\rho(z)=\rho_{z_i}=\sqrt{R_1^2-z_i^2},  \\&&{\rm radius\,\, of\,\, the\,\, shell\,\, for\,\, various\,\, z-coordinates,}\nonumber\\
 &&y_3=\sqrt{\rho_{z_i}^2-x_3},\\
 &&\sin{(\theta_i/2)}=\frac{y_3}{\rho_{z_i}},\quad\\&& {\rm angle\,\, of\,\, the\,\, circular\,\, segment.}\nonumber
\end{eqnarray}
The surface of the deformed spectator:
\begin{eqnarray}
      && S^{\rm iner.surf.}(b) = \sum_i  2 y_3 \Delta z,\\
      && S^{\rm circ.seg.}(b)= \frac{\rho^2_{z_i}}{2} (\theta_i-\sin{\theta_i}),\\
      && S^{s-c} (b)=\sum_i  (\rho_{z_i} +\rho_{z(i-1)})(1-\theta_i)\Delta z,\\
      && V^{s-c} (b)=\sum_i (\pi \rho^2_zi-S^{\rm iner.surf.}(b))\Delta z,
\end{eqnarray}
where: $S^{\rm circ.seg.}$ surface of circular segment, $S^{\rm iner.surf.}$ is the 
surface of contact plane between the sphere and the cylinder.

\item The radius of a reference sphere (after collapsing deformed spectator):
\begin{eqnarray}
     && R^{s-p} (b) = \bigg(\frac{3V^{\rm s-p}(b)}{4\pi}\bigg)^{1/3},\\
     && R^{s-s} (b) = \bigg(\frac{3V^{\rm s-s}(b)}{4\pi}\bigg)^{1/3},\\
     && R^{s-c} (b) = \bigg(\frac{3V^{\rm s-c}(b)}{4\pi}\bigg)^{1/3}.    
\end{eqnarray}
The surface of the reference spheres:
\begin{eqnarray}
     && S^{s-p}_{spect} (b) = \frac{4}{3}\pi (R^{s-p} (b))^{3},\\
     && S^{s-s}_{spect} (b) = \frac{4}{3}\pi (R^{s-s} (b))^{3},\\
     && S^{s-c}_{spect} (b) = \frac{4}{3}\pi (R^{s-c} (b))^{3}.    
\end{eqnarray}
\item The geometrical surface factor, which enters into the deformation energy formula (Eq.~(\ref{eqn.01})), 
repeating Eq.~(\ref{eqn.05a}):
\begin{eqnarray}
 B_{\rm surf.}(def) = \frac{S(def)}{S(sphere)}.
\end{eqnarray}
For the different scenarios considered:
\begin{eqnarray}
 && B_{\rm surf.}^{s-p}(b) = \frac{ S^{\rm s-p}(b)}{S^{s-p}_{spect} (b)},\\
  &&B_{\rm surf.}^{s-s}(b) = \frac{ S^{\rm s-s}(b)}{S^{s-s}_{spect} (b)},\\
  &&B_{\rm surf.}^{s-c}(b) = \frac{ S^{\rm s-c}(b)}{S^{s-c}_{spect} (b)}.\\
\end{eqnarray}
\item The radius of the new sphere, its surface and geometrical factor can be written for $i=p,s,c$ as:
\begin{eqnarray}
     && R^{s-i} (b) = \bigg(\frac{3V^{\rm s-i}(b)}{4\pi}\bigg)^{1/3},\\
     && S^{s-i}_{spect} (b) = \frac{4}{3}\pi (R^{s-i} (b))^{3},\\
     && B_{\rm surf.}^{s-i}(b) = \frac{ S^{\rm s-i}(b)}{S^{s-i}_{spect} (b)}.\\
\end{eqnarray}
\item Also the estimation of the mass and charge of the spectator for $i=p,s,c$ is calculated as:
\begin{eqnarray}
    && A^{s-i} (b) = A_{Pb}\frac{V^{\rm s-i}(b)}{V_{\rm sphere}},\\
   &&  Z^{s-i} (b) = Z_{Pb}\frac{V^{\rm s-i}(b)}{V_{\rm sphere}}.
\end{eqnarray}

\end{itemize}
%%%%%%%%%%%%%%%%%%%%%%%%%%%%%%%%%%%%%%%%%%%%%%%%%%%%%%%%%%%%%%%%%%%%%%%%%%%%%%
\section{Details of the Lublin-Strasbourg Drop model}\label{a2}
The macroscopic energy from the Lublin-Strasbourg Drop (LSD) model \cite{pomorski:2003,dudek:2004} used in this context has the following form for nucleus with the mass number $A$ and charge $Z$:
\begin{eqnarray}
   E_{\rm total}(A,Z;def)
   =
   E(A,Z)
   &+&
   E_{\rm Coul.}(A,Z;def)                                         \nonumber\\
   &+&
   E_{\rm surf.\,}(A,Z;def)                                         \nonumber\\
   &+&
   E_{\rm curv.}(A,Z;def),                                                                                
                                                                 \label{eqn.01}
\end{eqnarray}
where $def$ are the deformation parameters depending on the chosen 
parametrization of the shape. Above we find $E_{\rm Coul.}(A,Z;def)$ - 
deformation-dependent Coulomb electrostatic energy term, the surface, 
$E_{\rm surf.}(A,Z;def)$, and curvature, $E_{\rm curv.}(A,Z;def)$
terms. The first term on the right-hand side of Eq.\,(\ref{eqn.01}) 
denotes by definition the combined deformation-independent terms:
\begin{eqnarray}
      E(A,Z) 
      &=&  
      Z M_{\rm H} 
      + 
      (A-Z) M_{\rm n} 
      - 
      0.00001433 \, Z^{2.39}                                     \nonumber \\
      &+&
      E_{\rm vol.}(A,Z)
      +
      E_{\rm cong.}(A,Z),
                                                                 \label{eqn.02} 
\end{eqnarray}
where the term proportional to $Z^{2.39}$ is the binding energy of the
electrons whereas the other two terms represent $Z$ masses of the Hydrogen 
atom and $(A-Z)$ masses of the neutron,
respectively. The deformation-independent congruence energy term is $E_{\rm cong.}(A,Z)$.

The volume energy is parametrized as:
\begin{equation}
      E_{\rm vol.}(Z,A)
      =
      b_{\rm vol.}\,(1 - \kappa_{\rm vol.} \, I^2\,)\,A  ,      
                                                                 \label{eqn.03} 
\end{equation} 
where $I=(A-2Z)/(A+2Z)$ is introduced for brevity. All the parameters
appearing implicitly in Eq.\,(\ref{eqn.01}), such as $b_{\rm vol.}$=-15.4920~MeV and 
$\kappa_{\rm vol.}$=1.8601  and the ones that appear below, are taken from \cite{pomorski:2003}.

The Coulomb LDM term reads:
\begin{equation}
      E_{\rm Coul.}(A,Z;def)
      =
      \frac{3}{5} \, e^2 \frac{Z^2}{r_0^{ch} A^{1/3}}\, B_{\rm Coul.}(def) 
      - 
      C_{4}\frac{Z^2}{A}, 
                                                                 \label{eqn.04} 
\end{equation} 
with electric charge unit denoted as $e$, and the 
so-called charge radius parameter $r_0^{ch}$= 1.21725~fm, $C_{4}$=0.9181~MeV. The term proportional to
$Z^2/A$ represents the nuclear charge-density diffuseness-correction
whereas the deformation dependent term, $B_{\rm Coul.}(\alpha)$, denotes
the Coulomb energy of a deformed nucleus normalized to that of the
sphere with the same volume. 

The surface energy in the LDM form reads:  
\begin{equation}
      E_{\rm surf.}(A,Z;def)
      =
      b_{\rm surf.}\,(1 - \kappa_{\rm surf.} I^2\,)\,A^{2/3} 
      B_{\rm surf.}(def),
                                                                \label{eqn.05} 
\end{equation}
where $b_{\rm surf.}$ = 16.9707 MeV and $\kappa_{\rm surf.}$ =  2.2938.
The deformation dependent term is defined as the surface energy of a
deformed nucleus normalized to that of the sphere of the same volume:
\begin{equation}\label{eqn.05a}
 B_{\rm surf.}(def) = \frac{S(def)}{S(sphere)}.
\end{equation}

The curvature term is given by:
\begin{equation}
      E_{\rm curv.}(A,Z;def)
      =
      b_{\rm curv.}\,(1 - \kappa_{\rm curv.} \, I^2\,)\,A^{1/3} 
      B_{\rm curv.}(def)
                                                                \label{eqn.06} 
\end{equation}
with
\begin{equation}
     B_{\rm curv.}(def) 
     =
     \int_0^\pi \hspace{-2mm} d\vartheta \int_0^{2\pi} \hspace{-3mm} d\varphi
     \left[
          \frac{1}{R_1(\vartheta,\varphi;def)}
          +
          \frac{1}{R_2(\vartheta,\varphi;def)}
     \right], \qquad
                                                                \label{eqn.07} 
\end{equation}
where $R_1$ and $R_2$ are deformation-dependent principal radii of the
nuclear surface at the point-position defined by spherical angles $\vartheta$ and $\varphi$, $b_{\rm cur.}$= 3.8602~MeV and $\kappa_{\rm cur.}$= -2.3764  The LSD parameters have been fitted to all known experimental masses and well reproduced the fission barriers. 
%\end{appendix}
%\end{widetext}

\bibliographystyle{spphys}       % APS-like style for physics
\bibliography{cites_prc}

\end{document}